\documentclass[final,5p,times,twocolumn]{elsarticle}

\usepackage[utf8]{inputenc}
\usepackage[T1]{fontenc}
\usepackage{amsmath,amssymb}
\usepackage{graphicx}
\usepackage{booktabs}
\usepackage{xcolor}
\usepackage{hyperref}
\usepackage{array}
\usepackage{multirow}
\usepackage{tabularx}
\usepackage{rotating}
\usepackage{enumitem}
\usepackage{xurl}
\usepackage{pifont}  % gives \ding{} symbols (✓, ✗, etc.)
\usepackage{wasysym} % gives extra circle symbols like ◐ (wasysym provides \LEFTcircle, \CIRCLE, \RIGHTcircle, \Circle, etc.)

\newcolumntype{L}[1]{>{\raggedright\arraybackslash}p{#1}}
\newcolumntype{C}[1]{>{\centering\arraybackslash}p{#1}}

\journal{Computer Standards \& Interfaces}

\begin{document}

\begin{frontmatter}

\title{UGAF-ITS: A Standards Harmonization Framework and Validation Tool for Multi-Framework AI Governance in Distributed Intelligent Transportation Systems}

\author[hct]{Talal Ashraf Butt\corref{cor1}}
\ead{tbutt@hct.ac.ae}

\author[hct]{Muhammad Iqbal}
\ead{miqbal1@hct.ac.ae}

\author[cmu]{Razi Iqbal}
\ead{razi.iqbal@ieee.org}

\cortext[cor1]{Corresponding author}

\affiliation[hct]{organization={Higher Colleges of Technology},
            city={Fujairah},
            country={United Arab Emirates}}

\affiliation[cmu]{organization={Central Michigan University},
            city={Mount Pleasant},
            country={United States of America}}

\begin{abstract}
Organizations deploying AI-enabled Intelligent Transportation Systems face fragmented governance: ISO/IEC~42001 demands a certifiable management system, the EU AI Act imposes binding high-risk obligations from August~2026, and the NIST AI Risk Management Framework structures voluntary practice. Each instrument is internally coherent, yet they drive different control vocabularies, evidence expectations, and audit rhythms. In distributed ITS deployments where vehicle manufacturers, roadside integrators, and cloud operators each hold partial evidence and partial accountability, this fragmentation multiplies compliance effort and obscures incident traceability.
 
This paper introduces UGAF-ITS, a standards harmonization framework that consolidates 154 source obligations from the three instruments into 12 unified controls across eight governance domains through a reproducible five-phase crosswalk methodology. A three-tier operating model allocates each control to the vehicle, edge, or cloud tier where enforcement and defensible evidence production are feasible. An evidence backbone of 20 versioned artifacts supports a single audit package across all three frameworks without duplicating content.
 
We evaluate UGAF-ITS through an open-source governance engine applied to four architecturally distinct ITS deployment scenarios. The engine encodes the complete crosswalk catalog and executes eight compliance computations. Three-tier deployments achieve 91.7\% average scoped framework coverage with 45.9\% evidence reduction, complete bidirectional traceability, and 80\% of artifacts serving all three frameworks simultaneously. Partial deployments contract predictably, with reduction holding between 25\% and 50\% across archetypes. The siloed baseline is enumerated document by document and computed by the tool at run time. These results establish structural properties of the harmonized catalog under architectural variation. They do not establish operational compliance readiness, which external assessment with practitioners has yet to test. The tool, scenarios, and all reported results are publicly available for independent replication.
\end{abstract}


\end{frontmatter}

\section{Introduction}\label{sec:introduction}

Three governance instruments now compete for engineering attention whenever an organization builds, deploys, or operates an AI-enabled system. Certifiable management-system requirements arrive through ISO/IEC~42001~\cite{iso42001}. Binding high-risk obligations arrive through the EU AI Act, enforceable for systems entering the European market from August~2026~\cite{euaiact,euaiact_timeline}. Voluntary practice is structured across four lifecycle functions under the NIST AI Risk Management Framework~\cite{nistrmf}. Each instrument is internally coherent. Yet they drive different control vocabularies, different evidence expectations, and different audit rhythms. An organization that must satisfy all three faces a problem that no single instrument solves. Governance has to be implemented once and demonstrated three times, without maintaining parallel control libraries, parallel documentation packages, and parallel audit preparations.

This problem intensifies in distributed systems. Intelligent Transportation Systems distribute AI functions across vehicle perception stacks, roadside signal optimization, and cloud-based traffic coordination~\cite{sarker2020review,etsiits}. Those components execute at different tiers owned by different organizations, typically vehicle OEMs, infrastructure integrators, and cloud operators, and each organization holds partial evidence and partial accountability. A model update can originate in the cloud, alter behavior at an edge node, and surface risk in the vehicle. Three organizations then retain three slices of one evidence trail, and no party holds the complete record. Governance frameworks designed for single-organization contexts encounter friction under this structure~\cite{burton2017,isopas8800}.

Existing responses fall into three categories. Enterprise governance architectures propose unified control libraries mapped across standards and regulations, though they assume a single organizational boundary that does not match most ITS deployments~\cite{robles2025uaigf,hernandez2025open}. Compliance automation tools evaluate systems against individual regulations with increasing sophistication, without harmonizing those regulations into one operational vocabulary~\cite{sovrano2025,marisma2025}. Transportation-focused assurance work connects AI behavior to established safety practice, typically for one component or one standard, and without cross-framework traceability~\cite{isopas8800,burton2017,nelson2025}. Across the body of work surveyed in Section~\ref{sec:related_work}, none allocates governance obligations across vehicle, edge, and cloud boundaries under multi-party ownership.

The three instruments differ in legal force, in structure, and in the evidence they expect. Table~\ref{tab:framework_snapshot} summarizes those differences. What follows draws out the evidence requirements that become decisive once an ITS deployment spans several tiers and several owners.

\begin{table*}[!t]
\centering
\caption{Comparison of AI governance frameworks relevant to Intelligent Transportation Systems.}
\label{tab:framework_snapshot}
\footnotesize
\setlength{\tabcolsep}{5pt}
\renewcommand{\arraystretch}{1.35}
\begin{tabular}{@{} L{2.4cm} L{3.6cm} L{5.8cm} L{3.2cm} L{2.0cm} @{}}
\toprule
\textbf{Framework} &
\textbf{Structure} &
\textbf{Key Requirements} &
\textbf{ITS Relevance} &
\textbf{Compliance} \\
\midrule

\textbf{ISO/IEC 42001:2023} &
Clauses 4--10 (PDCA cycle) + 38 Annex~A controls across 9 control areas &
AI management system; risk assessment \& treatment; human oversight; data governance; continuous improvement &
Certifiable management system for OEMs and service providers &
Voluntary;\newline Certifiable \\
\midrule

\textbf{EU AI Act (2024/1689)} &
Articles 8--15 (high-risk) + Annexes III, IV, VI; $\sim$45 obligations &
Risk management system; technical documentation; transparency \& human oversight; accuracy \& robustness; cybersecurity &
ITS classified high-risk (Annex~III); mandatory for EU market &
Mandatory;\newline Aug~2026 \\
\midrule

\textbf{NIST AI RMF 1.0} &
4 functions: \textsc{govern}, \textsc{map}, \textsc{measure}, \textsc{manage}; 72 subcategories across 19 categories &
Trustworthy AI principles; context mapping \& risk measurement; continuous monitoring; incident response; stakeholder engagement &
Implementation guidance; U.S.\ federal alignment; international interoperability &
Voluntary;\newline Best practice \\

\bottomrule
\multicolumn{5}{@{}l}{%
  \parbox{\linewidth}{\vspace{3pt}\scriptsize
    \textbf{PDCA} = Plan-Do-Check-Act.\enspace
    High-risk AI under the EU AI Act includes vehicle safety components (Annex~III, \S2) and traffic management AI affecting road safety.\enspace
    The $\sim$45 EU AI Act obligations listed here reflect the full high-risk requirement set; the crosswalk extraction (Section~\ref{subsec:crosswalk_method}) scopes this to 37 implementable technical obligations after excluding conformity assessment body, market surveillance, and CE marking procedural items.
  }}\\
\end{tabular}
\end{table*}

\subsection{Three Instruments, Three Evidence Regimes}
\label{subsec:instruments}

Management-system structure comes first. An Annex~SL Plan-Do-Check-Act cycle under ISO/IEC~42001 connects governance fundamentals to 38 Annex~A controls across nine areas~\cite{iso42001}. In ITS deployments, where vehicle manufacturers, roadside integrators, and cloud operators hold different components and different logs, that structure supplies an auditable baseline without requiring centralized ownership. UGAF-ITS treats Annex~A as the primary control vocabulary and binds its controls to a finite evidence catalog through clause-level traceability links~\cite{iso23894}.

Legal force arrives with the EU AI Act, which applies from 2~August~2026~\cite{euaiact,euaiact_timeline}.\footnote{The Commission's November~2025 Digital Omnibus proposal links the application date for high-risk AI systems to the availability of harmonised standards, with the latest possible date being December~2027 for Annex~III systems. The underlying multi-framework compliance pressure is unchanged regardless of the final timeline.} Compliance becomes an engineering constraint under that regime, since design-time choices and operational evidence are both required, and not documentation assembled late in deployment. ITS deployments fall within the high-risk perimeter when they operate as safety components or support critical infrastructure management such as road traffic, and Articles~8 to~15 define the core obligations. In distributed settings, legal duties couple to where controls can be enforced and where evidence can be collected~\cite{stettinger2024assurance}.

Measurement depth comes from NIST AI RMF~1.0, which structures risk management into GOVERN, MAP, MEASURE, and MANAGE functions spanning the system lifecycle~\cite{nistrmf}. Vehicle systems, roadside infrastructure, and cloud services generate different risk signals observed in different administrative domains, which complicates threshold setting, incident triggers, and coordinated corrective action. UGAF-ITS draws on MEASURE and MANAGE to set evidence expectations for post-deployment monitoring, incident handling, and change control in multi-party architectures~\cite{chamola2023txai}.

Underneath the vocabulary differences, the three instruments converge on risk management processes, human oversight mechanisms, transparency provisions, and documentation practices. The compliance stakes differ sharply. The core concerns overlap substantially, and that overlap is what makes harmonization possible.

\subsection{ITS Architecture and Governance Context}
\label{subsec:its_context}

ITS deployments distribute AI functions across three tiers: vehicle, roadside, and cloud. Coordination occurs through V2X interfaces and standardized station architectures defined in ISO~21217, ETSI~EN~302~665, and ARC-IT~\cite{iso21217,etsi_en302665,arcit,yurtsever2020survey}. These standards reflect operational reality, since no single tier owns all decisions, all logs, or all change authority. A model update may originate in the cloud, alter behavior at the roadside, and surface risk in the vehicle, while each operator retains only partial visibility into the evidence trail.

UGAF-ITS uses the tier boundary as the primary unit of evidence responsibility. Vehicle components produce safety-relevant traces and intervention events. Roadside components produce decision logs and configuration baselines tied to site context. Cloud operations produce cross-site monitoring summaries, risk aggregation records, and controlled change documentation. This separation allows unified controls to attach where behavior executes and where verification is observable. Section~\ref{sec:mapping} examines the cases where that alignment holds and the cases where legal responsibility diverges from compute location.

\subsection{Governance Failures Under Multi-Framework Pressure}
\label{subsec:failures}

Section~\ref{subsec:instruments} and Section~\ref{subsec:its_context} described a fragmentation problem in narrative terms. Turning that narrative into something a tool can evaluate requires definitions rather than complaints. Five failure modes recur in multi-framework ITS compliance, and each admits a definition computable against a deployment description.

\begin{itemize}[leftmargin=1.2em,itemsep=2pt,topsep=3pt]
\item \textbf{F1, control duplication.} One governance intent is implemented as separate controls under each framework. The consolidation ratio in Section~\ref{subsec:results} measures how much of that duplication a unified catalog removes.
\item \textbf{F2, evidence duplication.} A single underlying record is produced several times in framework-specific packaging. Measurement compares the unified artifact count against a scope-aware siloed baseline, together with the number of frameworks each artifact serves.
\item \textbf{F3, trace discontinuity.} An obligation has no resolvable path to an artifact, or an artifact has no path back to an obligation. Exhaustive link traversal detects both directions.
\item \textbf{F4, ownership ambiguity.} The party responsible for an obligation cannot be determined from the deployment description alone. The share of activated controls carrying a determinate tier and owner assignment measures this directly.
\item \textbf{F5, cross-tier dependency invisibility.} Evidence chains cross organizational boundaries and appear in no single framework's audit trail. Detected chain counts make the coordination requirement visible.
\end{itemize}

These definitions convert a narrative complaint into quantities. Every metric reported in Section~\ref{sec:validation} traces back to one of the five.

\subsection{Research Questions}
\label{subsec:rqs}

Those five failures set the agenda. Three questions organize the work against them.

\begin{itemize}[leftmargin=1.2em,itemsep=2pt,topsep=3pt]
\item \textbf{RQ1.} To what extent can obligations drawn from ISO/IEC~42001, the EU AI Act, and the NIST AI RMF be consolidated into a single control set without losing clause-level traceability to their sources? This targets F1 and F3.
\item \textbf{RQ2.} Does allocating unified controls to architectural tiers yield a determinate accountability assignment when a deployment spans several organizations with divergent contractual positions? This targets F4.
\item \textbf{RQ3.} How do the structural properties of a harmonized catalog behave when deployment architecture varies, and do they degrade in ways a practitioner can interpret? This targets F2 and F5.
\end{itemize}

\begin{figure*}[!t]
\centering
\includegraphics[width=\textwidth]{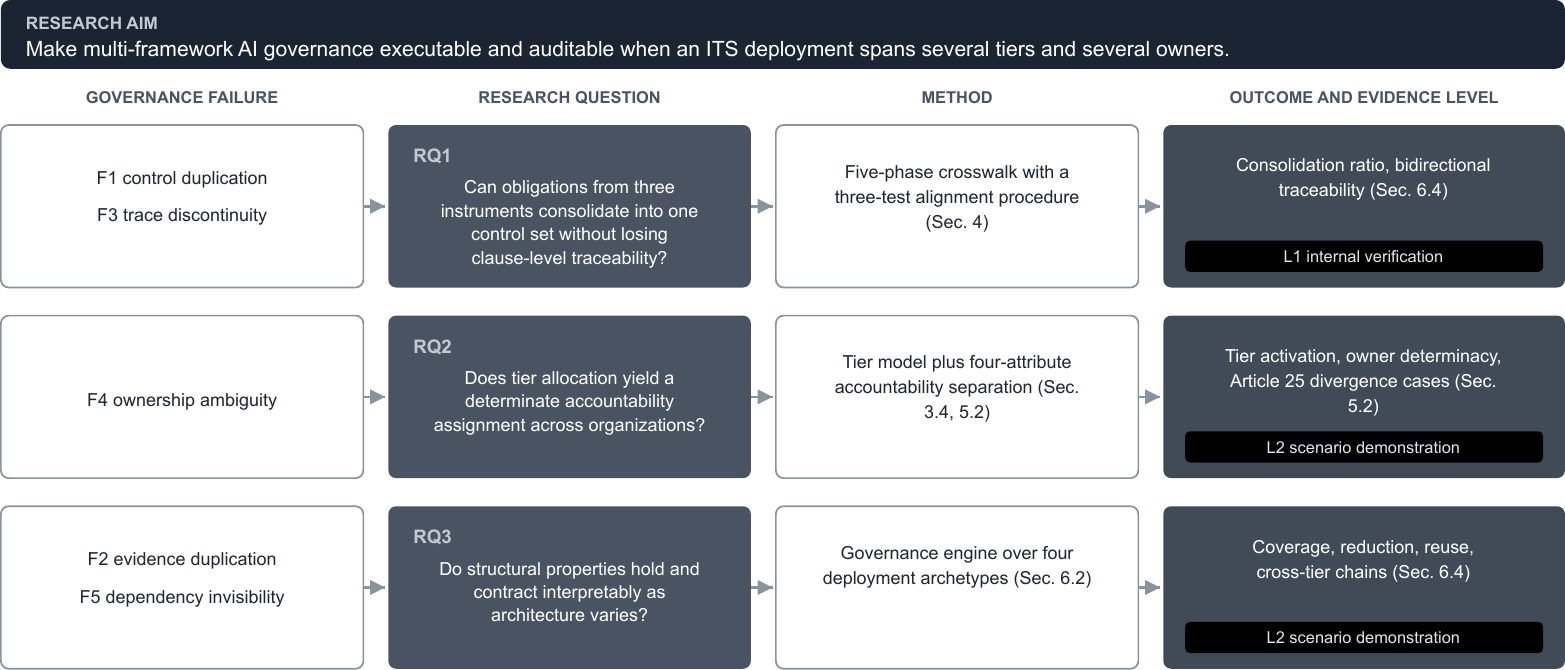}
\caption{Research design. Each governance failure defined in Section~\ref{subsec:failures} motivates one research question, which is addressed by a stated method and evaluated at a declared level of evidence. Level~L3, external validation, is outside the scope of this paper.}
\label{fig:research_design}
\end{figure*}

The crosswalk methodology in Section~\ref{sec:crosswalk} addresses RQ1, evaluated through the consolidation and traceability results in Section~\ref{subsec:results}. RQ2 is addressed by the tier model in Section~\ref{subsec:its_layer}, the component-level allocation in Section~\ref{sec:mapping}, and the accountability analysis in Section~\ref{subsec:accountability}. Multi-scenario evaluation in Section~\ref{sec:validation} addresses RQ3. Fig.~\ref{fig:research_design} sets out the full chain from failure definition through question and method to the level of evidence each result reaches.

\subsection{Contributions and Structure}
\label{subsec:contributions}

The primary contribution of this paper is a tier-aware governance allocation model. Under that model, a governance obligation attaches to the architectural tier where the control can be enforced and the supporting evidence can be observed. Twelve unified controls are assigned to the vehicle, edge/RSU, or cloud tier on that basis, so evidence duties follow execution and not organizational convenience. Section~\ref{sec:mapping} works through the allocation and treats the cases where legal responsibility, operational control, and compute location fail to coincide.

Three further contributions support that model. The crosswalk methodology supplies its input. A five-phase pipeline consolidates 154 source obligations, comprising 55 items from ISO/IEC~42001 (38 Annex~A controls plus 17 management-system clauses), 37 high-risk requirements from Articles~8--15 and Annex~IV of the EU AI Act, and 62 NIST AI RMF subcategories, into 12 unified controls under eight governance domains with clause-level bidirectional traceability preserved throughout.

The evidence backbone is the model's consequence. Twenty versioned artifacts support one audit package across all three frameworks, so a single record answers obligations that siloed compliance answers several times over.

The governance engine is the instrument of demonstration. An open-source tool encodes the complete catalog and executes eight compliance computations across four architecturally distinct deployment scenarios. The results establish structural properties of the framework under architectural variation. They do not establish operational compliance readiness, and Section~\ref{subsec:results} states that boundary before reporting any number.

The remainder of the paper is organized as follows. Section~\ref{sec:related_work} surveys existing governance approaches and identifies the research gap. Section~\ref{sec:architecture} presents the UGAF-ITS framework design, covering the harmonization layer, governance domains, tier model, and evidence backbone. Section~\ref{sec:crosswalk} details the crosswalk methodology and the unified control catalog. Section~\ref{sec:mapping} maps unified controls to representative ITS components and treats accountability under multi-party ownership. Section~\ref{sec:validation} states the evaluation design, describes the governance engine, and reports multi-scenario results. Section~\ref{sec:discussion} discusses implications, limitations, and directions for future work. Section~\ref{sec:conclusion} concludes.

\section{Related Work}
\label{sec:related_work}

AI governance research relevant to ITS spans enterprise control libraries, compliance automation, and transportation-specific assurance. Each stream has produced useful results. The evidence base is uneven across those streams, and the unevenness is itself part of the problem this paper addresses. Enterprise harmonization work currently sits largely in practitioner publications and preprints and not in archival venues, so this section marks the status of each such source where it appears and rests comparative claims on peer-reviewed work where an equivalent exists. None addresses the problem that arises when vehicle manufacturers, roadside integrators, and cloud providers each hold part of the evidence trail and must demonstrate compliance against multiple frameworks simultaneously. The following subsections examine representative approaches before identifying the gap that UGAF-ITS addresses.

\subsection{Cross-Framework Governance Architectures}
\label{subsec:rw_governance}

Enterprise AI governance work has recently converged on a pragmatic idea: organizations need a compact control library that can be mapped to multiple frameworks and then used to drive reusable evidence and repeatable oversight. Forrester's AEGIS, an industry analyst report, positions a cross-referenced control set as a way to reduce governance load by aligning one library to multiple sources, including ISO/IEC 42001 and AI RMF as universal anchors with partial EU AI Act coverage~\cite{aegis2025}. The Unified Control Framework (UCF), released as a preprint, proposes a set of controls and validates the mapping against a specific regulation~\cite{ucf2025}. Guidance documents from CSA and ISACA reinforce the same pairing logic and encourage organizations to treat standards and regulations as mutually supportive and not competing compliance tracks~\cite{csa2025}. Robles and Mallinson~\cite{robles2025uaigf} propose a Unified AI Governance Framework through systematic review of over 100 existing frameworks, but the contribution is entirely theoretical: no software prototype, no domain application, and no quantitative validation.

Ontology and knowledge graph approaches offer a different path to harmonization. Golpayegani~et~al.\ map EU AI Act risk requirements to ISO~31000 using OWL2 ontologies, validated through SPARQL queries on the AIAAIC repository~\cite{golpayegani2022airo}. Hernandez~et~al.\ extend this with an open knowledge graph mapping EU AI Act concepts to ISO management system standards~\cite{hernandez2025open}. Neither includes the NIST AI RMF, neither produces a working compliance tool, and neither targets a specific deployment domain.

Pipeline-oriented governance has also emerged. Our earlier work, GEAP~\cite{butt2026pipeline}, encodes governance policies as deterministic gate rules within AI development pipelines, enforcing five compliance checkpoints and emitting signed conformity bundles. GEAP demonstrates that governance evidence can be produced as a byproduct of engineering execution. However, GEAP assumes centralized organizational ownership. UGAF-ITS addresses the complementary problem: governance under distributed ownership, where no single organization controls the full execution path. The two frameworks share a regulatory foundation but differ along every operational dimension: unit of analysis, enforcement mechanism, evidence design, and domain validation.

These efforts provide useful control abstractions, but they assume an enterprise boundary that does not match most ITS deployments. They do not allocate control ownership across vehicle, roadside, and cloud tiers, and they do not specify where evidence is produced when multiple operators hold different parts of the audit trail. UGAF-ITS builds on the emerging control-library pattern, but adds a deployment-aware distribution model and an evidence backbone that preserves traceability while making artifact reuse feasible across stakeholders.

\subsection{Compliance Automation Tools}
\label{subsec:rw_tools}

A parallel stream of work automates compliance assessment without harmonizing governance structures. MARISMA~\cite{marisma2025} provides the most mature example in the standards-and-tools space: a cybersecurity risk management framework supported by eMARISMA, a cloud-based platform validated across multiple companies in different countries and sectors, aligned with ISO/IEC~27000 and 31000. Related work at the intersection of standards and tool-backed validation includes extensions of the MUD standard for modelling cyber-physical system behavior~\cite{mudcps2024}. MARISMA demonstrates what CSI-grade tool-backed validation looks like (automated risk analysis, multi-sector case studies, and quantitative metrics), but it addresses a single standards family in a single risk domain without harmonizing heterogeneous governance instruments.

PASTA~\cite{pasta2026} (preprint) uses large language models to evaluate AI systems against five policies simultaneously, achieving Spearman $\rho \geq 0.626$ correlation with expert judgments across a 12-participant user study. PASTA represents the strongest quantitative validation in the compliance automation space, but it treats regulations individually without consolidating them into a unified control vocabulary. Sovrano~et~al.~\cite{sovrano2025} validate AI-assisted EU AI Act technical documentation through comparison with three legal experts, achieving statistically significant agreement, and release an open-source replication package. Their contribution targets a single framework with a document-generation strategy and not cross-framework control consolidation. Neither PASTA nor Sovrano addresses distributed system architectures or tier-aware evidence allocation.

\subsection{Transportation AI Governance}
\label{subsec:rw_transport}

Transportation focused governance work is strongest where it connects AI behavior to safety and assurance practices already established in automotive engineering. Functional safety (ISO 26262), safety of the intended functionality (ISO 21448), and cybersecurity engineering (ISO/SAE 21434) provide rigorous assurance methods, but none integrates with AI-specific governance frameworks like ISO/IEC 42001 or the EU AI Act. ISO/PAS 8800 extends safety thinking to AI-enabled road vehicle functions and motivates lifecycle assurance for learning components~\cite{isopas8800}. This orientation complements existing safety practices, but it is not designed as a multi-framework compliance approach. It does not provide clause-level integration of ISO/IEC 42001, the EU AI Act, and AI RMF, nor does it address distributed evidence collection across vehicle, roadside infrastructure, and cloud operations. Conceptual proposals connecting responsible AI development to automotive contexts move the discussion toward governance, yet they typically remain focused on a single standard's stack and do not operationalize cross-framework traceability~\cite{nelson2025}.

Work applying AI RMF to autonomous vehicle components provides useful specialization, but it also illustrates the limits of component-level profiling for ITS-scale governance. Carlton and Eshete's traffic sign recognition profile translates AI RMF categories into a concrete use-case checklist, but it targets one perception capability and not a system-of-systems spanning V2X and traffic management~\cite{carlton2025}. Safety case arguments for machine learning in automated driving show how to structure evidence for assurance claims, but do not resolve regulatory fragmentation or distributed accountability across tiers~\cite{burton2017}.

\subsection{Research Gap}
\label{subsec:rw_gap}

Existing work leaves a practical gap where ITS engineering meets multi-framework compliance. Enterprise governance architectures provide control abstractions and mapping logic, but do not allocate control ownership across vehicle, roadside, and cloud components with different operators. Compliance automation tools evaluate against individual frameworks with growing rigour, but they do not produce a unified control vocabulary or tier-aware evidence allocation. Transportation-focused work provides safety and assurance practices, but it typically targets one component or one standard's stack, limiting traceability to regulatory obligations.

This fragmentation creates concrete operational risk. Consider an incident involving pedestrian detection failure at a smart intersection. The vehicle manufacturer points to roadside data quality. The roadside vendor points to cloud model updates. The cloud operator points to edge deployment configuration. Each party demonstrates compliance with their interpretation of one framework, but no single party holds complete evidence, and root cause analysis stalls in jurisdictional disputes. Corrective actions satisfying ISO/IEC 42001 may conflict with EU AI Act documentation requirements. Meanwhile, the August 2026 applicability deadline approaches. Fragmented compliance efforts also force teams to produce parallel evidence packages for each framework, unaware that well-structured artifacts could serve multiple obligations simultaneously.

Table~\ref{tab:related_work} summarizes the gap across seven columns: domain scope, coverage of each governance framework (ISO 42001, EU AI Act and NIST AI RMF), ITS tier applicability, crosswalk methodology, and validation approach. None of the approaches surveyed here combines all three frameworks across vehicle, edge, and cloud boundaries with a reproducible crosswalk methodology and a software tool. The comparison rests on a targeted review of enterprise governance architectures, compliance automation tools, and transportation assurance work rather than on a systematic review protocol, so it bounds a claim about this sample and not about the whole literature. UGAF-ITS addresses this gap through four mechanisms: a reproducible crosswalk consolidating 154 source obligations into 12 unified controls with clause-level traceability, tier-aware control allocation assigning responsibilities to vehicle, edge, and cloud boundaries, an evidence backbone enabling artifact reuse while preserving audit defensibility across all three frameworks, and an open-source governance engine that computes these properties across multiple deployment scenarios.

\begin{table*}[!t]
\centering
\caption{Comparison of AI governance frameworks and integration approaches
for Intelligent Transportation Systems.
Within the approaches surveyed here, UGAF-ITS is the only one combining
all three frameworks with tier-aware control allocation and tool-backed
structural evaluation for distributed ITS deployments. The comparison reflects
a targeted review and not a systematic review protocol.}
\label{tab:related_work}
\footnotesize
\setlength{\tabcolsep}{4pt}
\renewcommand{\arraystretch}{1.3}
\begin{tabular}{@{} L{1.8cm} L{3.0cm} L{1.8cm} c c c c c L{2.2cm} @{}}
\toprule
\textbf{Category} &
\textbf{Reference} &
\textbf{Domain} &
\textbf{\fontsize{6.5}{7.5}\selectfont ISO 42001} &
\textbf{\fontsize{6.5}{7.5}\selectfont EU AI Act} &
\textbf{\fontsize{6.5}{7.5}\selectfont NIST RMF} &
\textbf{\fontsize{6.5}{7.5}\selectfont ITS Tiers} &
\textbf{\fontsize{6.5}{7.5}\selectfont Crosswalk} &
\textbf{Validation} \\
\midrule

%--- General AI Governance ---
\multicolumn{9}{@{}l}{\textit{General AI Governance}} \\[2pt]
& Forrester AEGIS~\cite{aegis2025}    & General AI & \CIRCLE & \LEFTcircle & \CIRCLE & \Circle & \LEFTcircle & Conceptual \\[2pt]
& Credo AI UCF~\cite{ucf2025}         & General AI & \LEFTcircle & \CIRCLE & \CIRCLE & \Circle & \Circle & Colorado case \\[2pt]
& CSA/ISACA~\cite{csa2025}            & General AI & \CIRCLE & \CIRCLE & \CIRCLE & \Circle & \Circle & Guidance only \\
\midrule

%--- Automotive & AV Safety ---
\multicolumn{9}{@{}l}{\textit{Automotive \& AV Safety}} \\[2pt]
& ISO/PAS 8800~\cite{isopas8800}          & Automotive    & \Circle & \Circle & \Circle & \LEFTcircle & \Circle & Standard \\[2pt]
& Nelson \& Lin~\cite{nelson2025}         & Automotive    & \LEFTcircle & \Circle & \Circle & \LEFTcircle & \Circle & Conceptual \\[2pt]
& Carlton \& Eshete~\cite{carlton2025}    & AV Perception & \Circle & \Circle & \CIRCLE & \LEFTcircle & \Circle & Profile \\
\midrule

%--- Pipeline Governance ---
\multicolumn{9}{@{}l}{\textit{Pipeline Governance}} \\[2pt]
& GEAP~\cite{butt2026pipeline} & AI Pipelines & \CIRCLE & \CIRCLE & \CIRCLE & \Circle & Pipeline gates & Synthetic case \\
\midrule

%--- Compliance Automation ---
\multicolumn{9}{@{}l}{\textit{Compliance Automation}} \\[2pt]
& MARISMA~\cite{marisma2025}           & Cybersecurity & \Circle & \Circle     & \Circle     & \Circle & \Circle & Multi-sector tool \\[2pt]
& PASTA~\cite{pasta2026}               & General AI    & \Circle & \LEFTcircle & \LEFTcircle & \Circle & \Circle & Expert correlation \\[2pt]
& Sovrano et al.~\cite{sovrano2025}    & General AI    & \Circle & \CIRCLE     & \Circle     & \Circle & \Circle & Expert validation \\
\midrule

%--- Ontology / Knowledge Graph ---
\multicolumn{9}{@{}l}{\textit{Ontology / Knowledge Graph}} \\[2pt]
& AIRO~\cite{golpayegani2022airo}            & General AI & \Circle     & \CIRCLE     & \Circle     & \Circle & \LEFTcircle & SPARQL queries \\[2pt]
& Robles \& Mallinson~\cite{robles2025uaigf} & General AI & \LEFTcircle & \LEFTcircle & \LEFTcircle & \Circle & \Circle     & Theoretical \\
\midrule

%--- This Work ---
\multicolumn{9}{@{}l}{\textit{This Work}} \\[2pt]
& \textbf{UGAF-ITS} & \textbf{ITS} &
\textbf{\CIRCLE} & \textbf{\CIRCLE} & \textbf{\CIRCLE} &
\textbf{\CIRCLE} & \textbf{\CIRCLE} & \textbf{Case study} \\

\bottomrule
\multicolumn{9}{@{}l}{%
  \parbox{\linewidth}{\vspace{3pt}\scriptsize
    $\CIRCLE$ = fully addresses;\enspace
    $\LEFTcircle$ = partially addresses;\enspace
    $\Circle$ = not addressed.\\[2pt]
    AEGIS maps 39 controls but lacks ITS architecture.\enspace
    UCF proposes 42 controls without tier distribution.\enspace
    ISO/PAS~8800 targets functional safety, not governance harmonization.\enspace
    MARISMA validates via eMARISMA cloud tool across 4+ sectors (ISO~27000/31000).\enspace
    PASTA achieves Spearman $\rho \geq 0.626$ with expert judgments.\enspace
    AIRO maps EU~AI~Act to ISO~31000 via OWL2.\enspace
    None of the surveyed approaches combines all three frameworks with tool-backed evaluation for distributed ITS.
  }}\\
\end{tabular}
\end{table*}

\section{UGAF-ITS Framework Design}
\label{sec:architecture}

UGAF-ITS treats AI governance in ITS as an operational engineering system that must hold under distribution across deployment tiers and organizational owners. The framework compiles obligations from ISO/IEC~42001, the EU AI Act, and the NIST AI RMF into unified controls with explicit traceability to source identifiers, then assigns control ownership and evidence duties to the vehicle, edge and roadside unit (RSU), and cloud tiers.

\subsection{Framework Overview and Design Rules}
\label{subsec:arch_overview}

\begin{figure*}[!t]
\centering
\includegraphics[width=\textwidth]{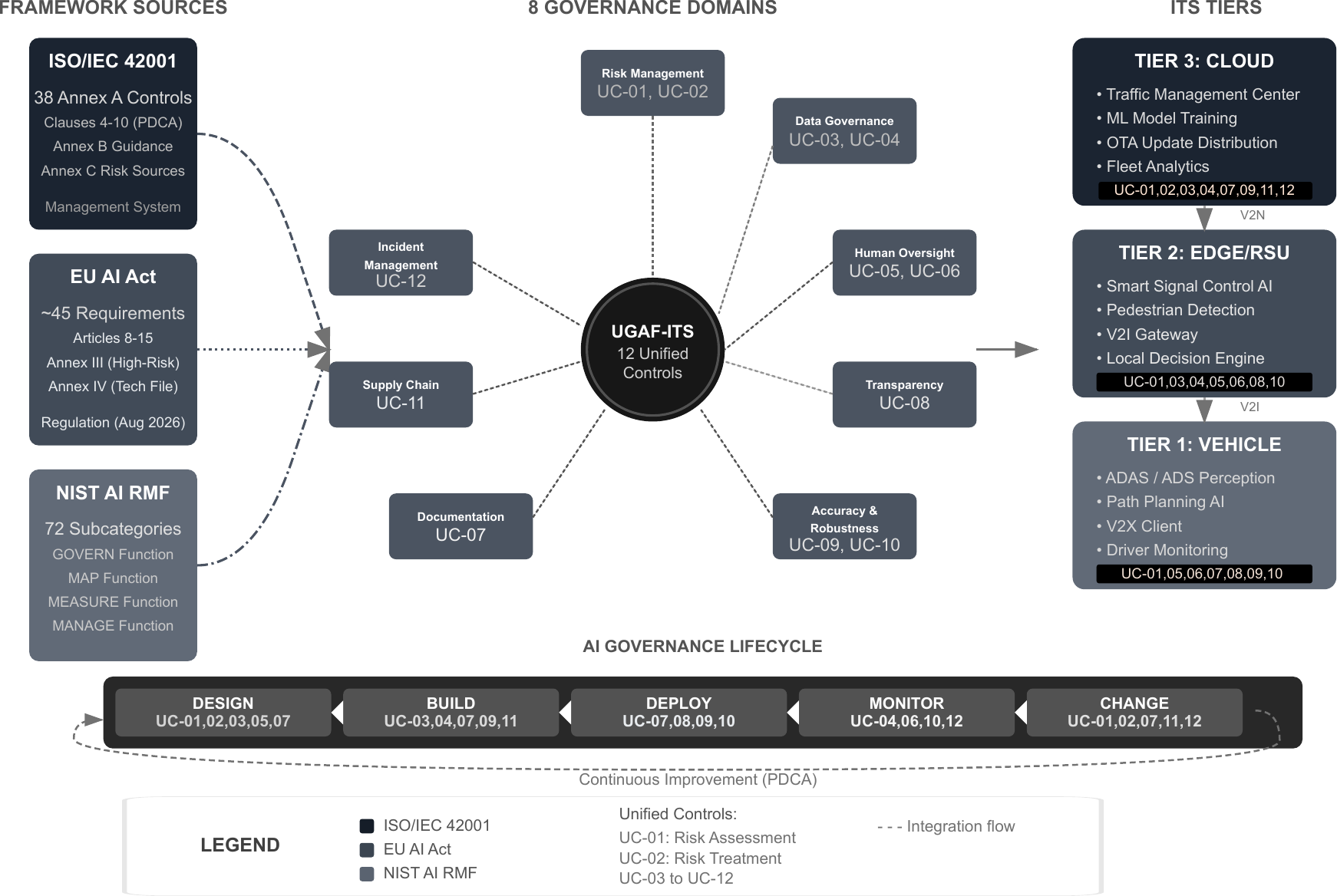}
\caption{UGAF-ITS framework integrating three framework sources (left) 
through eight governance domains (center) to ITS deployment tiers (right), 
governed across the AI lifecycle (bottom). The architecture consolidates 
154 source obligations into 12 unified controls mapped to vehicle, 
edge/RSU, and cloud tiers. \textit{Primary UCs shown per tier; see 
Table~\ref{tab:unified_controls} for complete applicability.}}
\label{fig:architecture}
\end{figure*}

ITS programs face two forms of fragmentation that break conventional compliance: engineering behavior executes across vehicle, roadside, and cloud components, while evidence ownership spans authorities, integrators, suppliers, and operators. UGAF-ITS addresses this fragmentation by converting heterogeneous framework items into a compact unified control catalog and binding each control to concrete evidence artifacts that can be produced where the relevant signals exist. Fig.~\ref{fig:architecture} summarizes the end-to-end flow from framework sources to unified controls, governance domains, tier-scoped artifacts, and lifecycle phases. The key constraint is that governance attaches to the tier where behavior executes and evidence can be observed, while preserving trace links back to the originating framework obligations.

UGAF-ITS follows four design rules that keep governance implementable and defensible under audit. Each unified control retains identifier-level links to ISO clauses and Annex~A controls, EU AI Act articles and annex items, and NIST AI RMF subcategories. Without this traceability, an auditor cannot recover the compliance rationale behind a unified control. Control ownership is tier aware: each control is assigned to the tier that can enforce it and observe evidence without inserting heavyweight checks into real-time loops. Because models, sensors, and operating policies change continuously, evidence artifacts evolve under version control instead of being reconstructed at release time. When obligations conflict across frameworks, tie-break decisions are recorded as lightweight Architecture Decision Records (ADRs) that capture precedence, scope boundaries, and tier placement rationale.

Unified controls specify what must hold, while the evidence backbone specifies what must exist to demonstrate that it holds. For example, risk assessment remains satisfied only when risk entries, monitoring thresholds, and incident triggers remain consistent with the deployed configuration and its operational context.

\subsection{Framework Harmonization Layer}
\label{subsec:framework_layer}

The framework harmonization layer compiles 154 source items into 12 unified controls organized across eight governance domains. Harmonization separates shared governance intent from framework specific extensions that must remain trace-linked for legal or procedural completeness. The resulting catalog (Table~\ref{tab:unified_controls}) preserves identifier-level traceability while avoiding framework wise reimplementation. As a representative alignment, ISO/IEC~42001 risk and continual-improvement requirements, EU AI Act risk-management obligations (e.g., Article~9), and NIST AI RMF risk functions align into UC-01/UC-02, while procedural duties that do not harmonize cleanly remain explicit extensions instead of being absorbed into a generic control. Section~\ref{sec:crosswalk} details the crosswalk pipeline and coding rules that produce this catalog.

\subsection{Governance Domains and Evidence Artifacts}
\label{subsec:arch_domains}

UGAF-ITS organizes implementation into eight governance domains that remain stable even as individual regulations evolve. Each domain acts as an ownership boundary that binds obligations to a primary evidence artifact, which reduces interpretation drift and gives auditors stable ``questions'' to ask across stakeholders and tiers. The eight domains (Risk Management (D1), Data Governance (D2), Human Oversight (D3), Transparency (D4), Accuracy \& Robustness (D5), Documentation (D6), Supply Chain (D7), and Incident Management (D8)) are defined by their framework anchors and unified controls in Table~\ref{tab:unified_controls}, and their consolidation properties are quantified in Table~\ref{tab:consolidation_domain}.

Two structural choices make the domain layer maintainable. Each domain anchors to a primary artifact that can be versioned, reviewed, and defended under audit questioning, instead of distributing compliance across clause-level narrative. Artifacts preserve bidirectional traceability from source identifiers to unified controls and from unified controls to evidence, enabling reuse without eroding accountability.

The Technical File serves as the glue artifact. It connects requirements, design decisions, validation evidence, deployment records, and change notices into an assessor navigable record.

Domain placement follows evidence feasibility. Artifacts requiring aggregation and governance decisions concentrate in cloud operations, while artifacts requiring local observability concentrate at the edge. Vehicle tier artifacts prioritize safety relevant constraints and cues facing operators, while exporting trace identifiers and version hashes that link into shared records. For V2X-relevant interaction points, vehicle-tier clients that consume Signal Phase and Timing (SPaT) and MAP messages primarily evidence UC-05/UC-06 through understandable status cues and safe interruption behavior, while edge/RSU components evidence UC-03/UC-04 through sensor-health gates, calibration baselines, and disparity checks tied to the site configuration.

\subsection{Three-Tier ITS Governance Model}
\label{subsec:its_layer}

\begin{figure*}[!t]
\centering
\includegraphics[width=\textwidth]{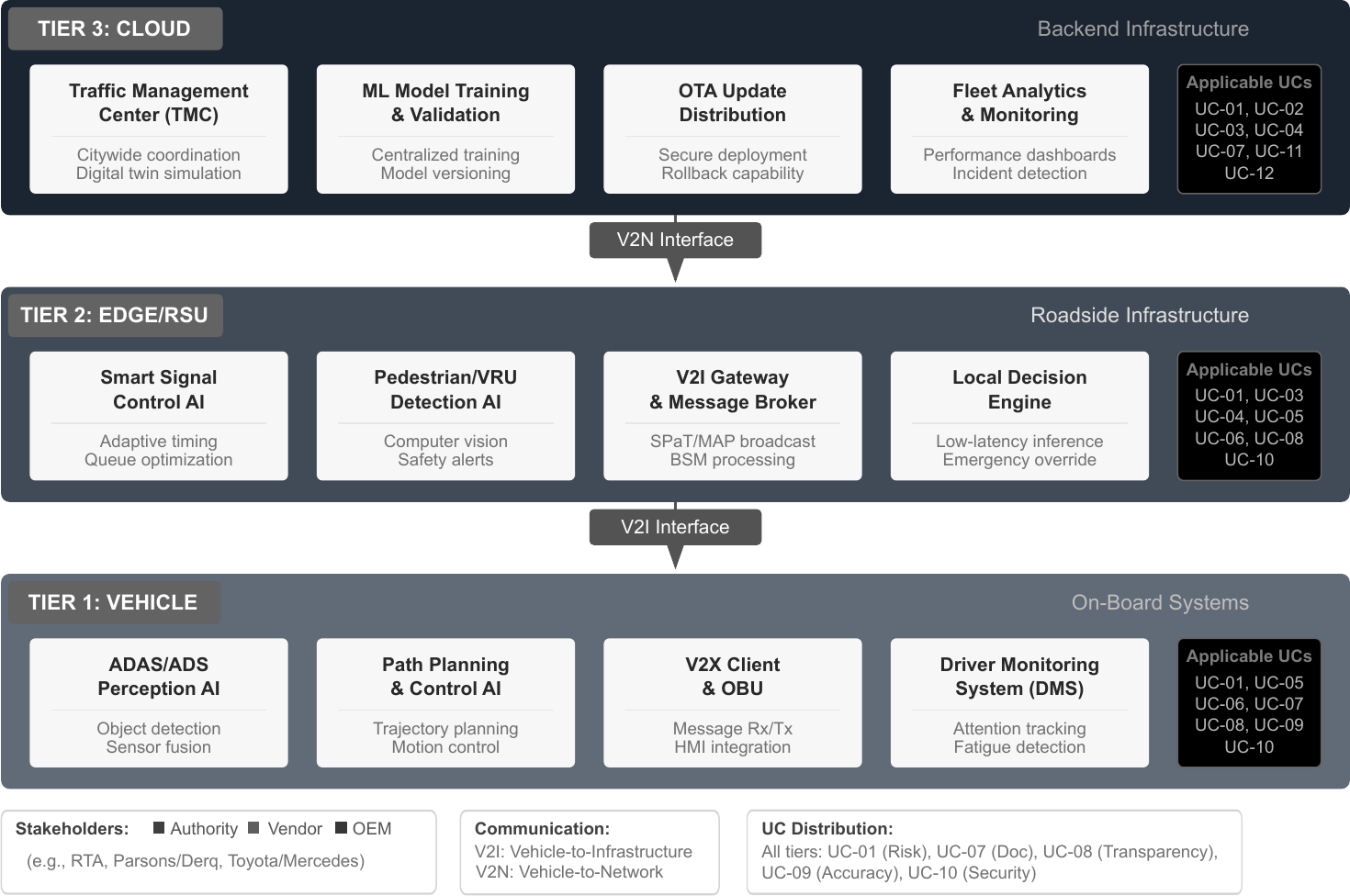}
\caption{Three-tier ITS governance model with AI components and 
unified control (UC) applicability at each tier. Vehicle tier addresses 
on-board AI under real-time safety constraints; edge/RSU tier encompasses 
roadside AI with multi-stakeholder coordination; cloud tier supports 
centralized operations requiring governance across all eight domains. \textit{Primary 
UCs highlighted; complete tier applicability in Table~\ref{tab:unified_controls}.}}
\label{fig:three_tier}
\end{figure*}

Fig.~\ref{fig:three_tier} assigns governance duties to the tier that can enforce the control, observe and retain defensible evidence, satisfy latency and safety constraints, and align accountability with operational ownership. The model treats vehicle, edge/RSU, and cloud as governance boundaries and not as mere compute placement. Evidence responsibilities then remain determinate when vendors and operators differ by tier. Legal responsibility is a separate attribute that the architecture does not fix, and Section~\ref{subsec:accountability} sets out how the two are reconciled when they diverge.

\textbf{Vehicle tier.} On-board AI executes under strict real-time and safety constraints. Governance emphasizes operator interaction, intervention paths, and deterministic traces and not heavy online checks. Evidence at this tier includes override and fallback events, safety-relevant logs scoped to operational design domains, validation summaries, and security test evidence for deployed binaries.

\textbf{Edge/RSU tier.} Roadside intelligence influences traffic behavior through local sensing, perception, and actuation, often across organizational boundaries. Governance emphasizes local context preservation and defensible escalation: decision logs, configuration baselines, calibration states, sensor-health gates, and incident triggers.

\textbf{Cloud tier.} Central operations aggregate cross-site signals and govern releases and suppliers. Cloud evidence concentrates on risk management, documentation integrity, and change governance, providing the strongest locus for cross-tier traceability and audit packaging.

Section~\ref{sec:mapping} specifies the detailed control-to-component allocation and evidence sources for each tier.

\subsection{Evidence Backbone}
\label{subsec:evidence}

UGAF-ITS uses an evidence backbone to make multi-framework compliance executable in distributed ITS programs~\cite{naja2022kg}. The backbone defines a finite artifact set that can be produced once and traced to obligations across multiple standards. This structure reduces duplication without weakening audit readiness because each artifact retains explicit trace links, tier ownership metadata, and configuration binding.

Under the scoped instantiation, the evidence backbone comprises 20 artifacts that jointly satisfy the three frameworks, compared with the 37 documents enumerated in \ref{app:register} for a siloed approach. The consolidation yields a 45.9\% reduction in documentation volume while preserving traceability, and each artifact supports 2.8 frameworks on average. Most artifacts serve all three frameworks because they capture shared governance intent such as risk management, documentation, monitoring, and incident response.

Evidence reuse remains bounded by audit constraints. Artifacts must remain configuration-tied under version control~\cite{kreuzberger2023mlops}, preserve tier context without collapsing vehicle and roadside evidence into cloud narratives, and answer stable audit questions with pointers to supporting records.

In multi-owner deployments, the backbone assumes evidence exchange contracts and not raw log centralization. Minimum exchange objects include signed incident summaries, artifact hashes, configuration baselines, and escalation records. Raw traces remain locally retained under defined access conditions. This model supports incident-to-change closure and technical file integrity without requiring vertical integration.

\section{Framework Crosswalk Methodology}
\label{sec:crosswalk}

UGAF-ITS relies on a reproducible crosswalk that transforms heterogeneous governance sources into a compact, traceable control catalog suitable for distributed ITS deployments. The crosswalk takes atomic obligations from ISO/IEC~42001, the EU AI Act, and the NIST AI RMF, aligns them through a structured coding protocol, and synthesizes unified controls that retain links back to clauses, articles, and subcategories. It then maps each unified control to evidence artifacts at each tier, so stakeholders can implement controls where enforcement and observability exist.

\subsection{Crosswalk Pipeline}
\label{subsec:crosswalk_method}

\begin{figure*}[!t]
\centering
\includegraphics[width=\textwidth]{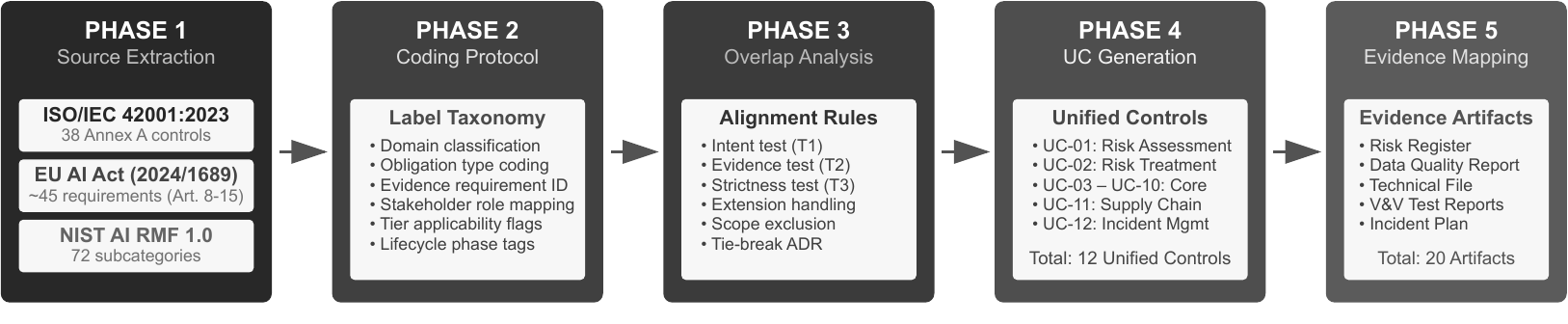}
\caption{Five-phase crosswalk methodology for harmonizing AI governance frameworks into UGAF-ITS unified controls. The process transforms 154 source obligations through systematic coding, overlap analysis, and UC synthesis to produce 12 unified controls and 20 evidence artifacts.}
\label{fig:crosswalk}
\end{figure*}

Fig.~\ref{fig:crosswalk} summarizes the five-phase pipeline used to derive the unified control catalog and the associated evidence backbone.

\textbf{Phase~1: Source extraction.} We decomposed each framework into atomic obligations that can be assessed independently and mapped without losing intent. Each atom was recorded with its identifier, a one-sentence obligation statement, and the minimal evidence implication needed to demonstrate execution in practice. The extraction yielded 55 items from ISO/IEC~42001 (38~Annex~A controls plus 17 management-system clauses from Clauses~4--10), 37 items from the EU AI Act, and 62 items from the NIST AI RMF, totalling 154 source obligations.

\emph{Scoping decisions.} Not all framework items entered the crosswalk. The EU AI Act contains approximately 45 high-risk obligations across Articles~8--15 and Annex~IV, but eight of those target conformity assessment bodies, market surveillance authorities, and CE marking procedures. These are regulatory workflow obligations. They execute outside the engineering scope and cannot be operationalized as technical controls, so we excluded them. That leaves 37 items. The NIST AI RMF defines 72 subcategories across 19 categories. Ten of those address organizational profiling (GOVERN~1, GOVERN~2) and stakeholder engagement, activities that set context for governance but precede control instantiation. Excluding them yields 62 items. ISO/IEC~42001 moved in the opposite direction. We started with 38 Annex~A controls and added 17 management-system clauses from Clauses~4--10 that impose implementable requirements: risk assessment planning, documented information, operational control, and performance evaluation. That gives 55 items. The three extractions total 154 source obligations.

\textbf{Phase~2: Coding protocol.} We applied a structured label taxonomy to each atom to enable consistent alignment across frameworks. The taxonomy assigns (i)~governance domain (D1--D8), (ii)~obligation type (preventive, monitoring, documentation, oversight, or incident control), and (iii)~evidence requirement class (policy, register, report, log, plan, or technical file). It also captures stakeholder role, tier applicability (T1~vehicle, T2~edge/RSU, T3~cloud), and lifecycle phase (design, build, deploy, monitor, change). These labels are not narrative annotations; they are the matching keys used in the overlap analysis.

\textbf{Phase~3: Overlap analysis and alignment rules.} Alignment decisions run through three tests applied to every candidate pair of atoms, and the outcome pattern determines the classification without further judgment. Table~\ref{tab:decision_rules} gives the full procedure.

The intent test (T1) asks whether both atoms name the same governance object and require the same action on it. Governance objects are drawn from a closed list fixed in Phase~2, covering the risk register, datasets, oversight mechanisms, the documentation set, suppliers, and incidents. Two atoms that constrain different objects never merge, however similar their wording.

The evidence test (T2) asks whether one artifact, produced once, would satisfy an assessor querying either atom. Failure here does not block a merge, because a unified control states what must hold, not which document proves it. Failure instead forces the evidence backbone to carry two artifact bindings for the same control.

The strictness test (T3) asks whether every implementation satisfying one atom also satisfies the other. Strictness is a partial order and not a ranking, so three outcomes are possible. Symmetric satisfaction means the atoms are interchangeable. Asymmetric satisfaction means one atom dominates and supplies the control objective. Incomparable atoms constrain different aspects of the same action, and the control objective then takes their conjunction with both retained as trace links.

Atoms failing T1 divide by actor. An atom constraining the same governance object through a different action becomes a framework-specific extension, trace-linked to the nearest unified control without merging into it. An atom whose action can be discharged only by an external body, such as a notified body or a market surveillance authority, leaves the crosswalk at Phase~1 scoping and reappears as a classified coverage gap in Section~\ref{subsec:results} rather than disappearing.

\begin{table}[!t]
\centering
\caption{Alignment decision procedure. Every candidate atom pair resolves to exactly one classification. T1 = intent test, T2 = evidence test, T3 = strictness test.}
\label{tab:decision_rules}
\footnotesize
\renewcommand{\arraystretch}{1.15}
\begin{tabular}{@{} c c c L{2.5cm} L{2.2cm} @{}}
\toprule
\textbf{T1} & \textbf{T2} & \textbf{T3} & \textbf{Classification and action} & \textbf{Instance} \\
\midrule
\checkmark & \checkmark & sym. & Full alignment. One control, one artifact binding & UC-01 risk register \\
\checkmark & \checkmark & asym. & Dominant strictness. Control takes the stronger objective; weaker atom trace-linked; ADR recorded & UC-07, Annex~IV dominant \\
\checkmark & $\times$ & any & Evidence-divergent overlap. One control, several artifact bindings & UC-03 data quality \\
\checkmark & any & incomp. & Conjunctive merge. Control takes the conjunction; both atoms trace-linked & UC-05, UC-06 oversight \\
$\times$ & same obj. & n/a & Framework-specific extension. Trace-linked to nearest control, no merge & Annex~IV sub-items \\
$\times$ & ext. actor & n/a & External actor. Excluded at Phase~1, reported as gap class & Conformity assessment \\
\bottomrule
\end{tabular}
\end{table}

\textbf{Phase~4: Unified control synthesis and tie-breaks.} Aligned atoms group into a single control objective carrying explicit trace links to every contributing source requirement. Conflicts resolve to the strictest implementable requirement, and the phrase carries a precise meaning. An atom is stricter when its satisfaction set is a proper subset of the alternative's, so any implementation meeting it also meets the alternative. An atom is implementable when a party inside the deployment can discharge it through engineering or documented organizational action, which excludes obligations requiring an external body to act. Weaker atoms are retained as trace-linked subsets and not dropped. In practice, legal obligations constrain the minimum baseline, standards supply auditable management-system structure, and voluntary guidance strengthens measurement depth.

Incomparable conflicts receive different treatment, and they are the reason the catalog preserves framework-specific extensions at all. Human oversight illustrates the pattern. Article~14 specifies intervention mechanisms in operational detail, ISO/IEC~42001 Annex~A.15 demands documented management-system treatment of the same function, and neither satisfaction set contains the other. No dominant atom exists, so UC-05 and UC-06 take the conjunction and both sources remain trace-linked. Human oversight consequently consolidates less aggressively than any other multi-control domain, a result visible in Section~\ref{subsec:results}.

Granularity follows from this procedure rather than from a target. One unified control exists for each pairing of governance object and required action that survives T1 clustering, which is why data governance yields two controls while documentation yields one. Twelve is an output of the rule set applied to 154 atoms, not a number chosen in advance. Section~\ref{subsec:granularity} tests what happens to the reported metrics when the clustering threshold is deliberately loosened and tightened.

A concrete tie-break illustrates the process. UC-01 (AI Risk Assessment) aligns ISO/IEC~42001 Clause~6.1 (``actions to address risks''), EU AI Act Article~9(2a--c) (``continuous risk management system''), and NIST AI RMF MAP~5.1--5.2 (``risk identification and analysis''). The EU AI Act mandates that risk management be ``continuous'' and ``iterative throughout the entire lifecycle,'' which is stricter than ISO~42001's PDCA-based review cycle and NIST's voluntary guidance. The ADR records: \emph{precedence}~= EU AI Act Art.~9 (mandatory, strictest); \emph{scope}~= lifecycle-continuous risk register; \emph{tier placement}~= all tiers (T1--T3), anchored at cloud for aggregation. The weaker obligations remain trace-linked so that ISO~42001 auditors and AI RMF practitioners can recover their framework-specific rationale from the unified control.

A different conflict pattern appears in UC-07 (System Documentation). ISO/IEC~42001 Clause~7.5 asks for general documented information. NIST AI RMF GOVERN~1.3--1.5 and MEASURE~2.1 ask for risk governance and measurement documentation. Neither specifies the structure or content of those documents in detail. The EU AI Act does. Article~11 and Annex~IV prescribe exactly what the technical documentation must contain: training methodologies, data sets used, evaluation procedures, design choices, and more. The conflict here is not about lifecycle continuity (as in UC-01) but about \emph{documentation specificity}. The ADR records: \emph{precedence}~= EU AI Act Annex~IV (mandatory, most granular content requirements); \emph{scope}~= Annex~IV-structured Technical File with sections that also satisfy ISO and NIST requirements; \emph{tier placement}~= all tiers (T1--T3), maintained centrally at cloud. A single UC at this abstraction level keeps the control vocabulary compact. In practice, implementers should use Annex~IV as the minimum content template and map ISO~42001 and NIST trace links to the corresponding sections within it.

\textbf{Phase~5: Evidence mapping.} We mapped each unified control to a primary evidence artifact and to the tiers that can produce and maintain that evidence reliably. Evidence mapping favors reuse by assigning artifacts that satisfy multiple unified controls through structured sections and cross-references, while retaining version and ownership metadata for audit defensibility.

The method produces three outputs: a set of 12 unified controls that consolidate 154 source obligations, a trace map that records how each unified control links to specific source requirements, and an evidence mapping that assigns artifacts and tier applicability for implementation and audit use. Reproducibility is supported through an explicit codebook, recorded tie-break rules, and a decision log that preserves why alignments were accepted as full, partial, or out-of-scope. This structure reduces semantic drift when frameworks evolve and enables independent replication of the mapping from source identifiers to unified controls.

Two complementary validation steps strengthen confidence in the crosswalk outputs. Inter-rater agreement on a sampled subset of requirements under the Phase-2 coding protocol is planned as follow-on work. In the current version, the open-source governance engine described in Section~\ref{sec:validation} provides a computational check: it encodes all 154 obligations, traverses every traceability link exhaustively, and flags any broken or orphaned mappings. All 154 obligations resolve to at least one unified control, and all 12 unified controls trace back to obligations from all three source frameworks.

\subsection{Unified Control Catalog}
\label{subsec:uc_catalog}

Table~\ref{tab:unified_controls} presents the UGAF-ITS unified control catalog, comprising 12 controls that harmonize 154 source obligations across ISO/IEC~42001, the EU AI Act, and the NIST AI RMF. Each UC is defined by a control objective and a trace map that links it back to the originating clause, article, or subcategory identifiers. This traceability allows an implementer to start from a UC and recover the exact compliance rationale required by a regulator, auditor, or partner, without maintaining separate, framework-specific control libraries.

The table is designed to support implementation decisions and not to serve only as a mapping artifact. The framework columns show which sources contribute to each UC, and the tier column indicates where the control can be enforced and where evidence can be observed reliably. Controls that depend on operational signals or real-time interaction are emphasized in the vehicle and edge tiers, while controls that require aggregation and version governance extend into the cloud. UC-03 (Data Quality Assurance) and UC-04 (Bias Detection \& Mitigation) are primarily implemented in the edge and cloud tiers where data pipelines and training workflows reside. UC-05 (Human Oversight Design) and UC-06 (Override \& Intervention) are critical at the vehicle and edge tiers, where human-AI interaction and safe interruption must be possible under timing constraints. Cross-cutting controls such as UC-07 (System Documentation), UC-09 (Accuracy Validation), and UC-10 (Robustness \& Security) span all tiers because they bind system behavior to auditable evidence across the lifecycle.

UC-07 deserves scrutiny because it consolidates 18 source obligations into a single control. The three frameworks ask for documentation at very different levels of specificity. ISO/IEC~42001 Clause~7.5 requires ``documented information'' in general terms. NIST AI RMF GOVERN~1.3--1.5 asks for governance documentation without prescribing content structure. The EU AI Act is different. Annex~IV lists exactly what the technical documentation must contain: training methodologies, data sets, evaluation procedures, design choices, and more. A single UC at this abstraction level keeps the control vocabulary compact, but implementers need to understand that Annex~IV sets the floor. The UC-07 tie-break (Section~\ref{sec:crosswalk}) records this precedence explicitly, and the evidence backbone operationalizes it through a Technical File structured around Annex~IV content requirements with ISO and NIST trace links mapped to corresponding sections.

A concrete mapping example illustrates how the catalog preserves intent while enabling reuse. UC-01 (AI Risk Assessment) aligns ISO/IEC~42001 Clause~6.1 and Annex~A risk-focused controls with the EU AI Act obligation to maintain a continuous risk management system, and with the NIST AI RMF's MAP function practices for risk identification and analysis. In an ITS deployment, this unified control becomes executable when the risk register records hazards and misuse conditions across vehicle, roadside, and cloud components; and when monitoring thresholds and incident triggers are derived from those risks. The same risk register entry set then serves multiple purposes: it supports management-system audit checks under ISO/IEC~42001, it provides structured evidence of risk management under the EU AI Act, and it operationalizes AI RMF mapping and measurement decisions. The UC therefore functions as the contract, while the evidence artifacts provide the auditable proof that the contract is being executed in the deployed system.

\begin{table*}[!t]
\centering
\caption{UGAF-ITS unified control catalog with cross-framework traceability.
Twelve unified controls (UC-01--UC-12) are organized under eight governance
domains; each control traces to specific clauses across all three source
frameworks. Tier activation
($\blacksquare$\,=\,active, $\square$\,=\,not applicable)
shows where each control is enforced across the
vehicle--edge--cloud architecture.}
\label{tab:unified_controls}
\footnotesize
\setlength{\tabcolsep}{3.5pt}
\renewcommand{\arraystretch}{1.25}
\begin{tabular}{@{} L{1.6cm} c L{2.0cm} L{4.2cm} L{2.0cm} L{1.6cm} L{2.0cm} c c c @{}}
\toprule
\textbf{Domain} &
\textbf{UC} &
\textbf{Control Name} &
\textbf{Control Objective} &
\textbf{ISO 42001} &
\textbf{EU AI Act} &
\textbf{NIST RMF} &
\textbf{T1} & \textbf{T2} & \textbf{T3} \\
\midrule

% D1 — Risk Management
\multirow{2}{*}{\parbox{1.6cm}{\raggedright D1 Risk\newline Management}}
  & 01 & AI Risk Assessment
  & Identify and evaluate AI-specific risks incl.\ safety, bias, and security
  & Cl.\,6.1;\newline A.4.1--4.3
  & Art.\,9(2a--c)
  & \textsc{map}\,5.1--5.2
  & $\blacksquare$ & $\blacksquare$ & $\blacksquare$ \\[2pt]

  & 02 & Risk Treatment
  & Implement controls to mitigate identified risks to acceptable levels
  & Cl.\,8.3;\newline A.5.1--5.2
  & Art.\,9(2d),\newline 9(5)
  & \textsc{mg}\,1.2--1.4
  & $\blacksquare$ & $\blacksquare$ & $\blacksquare$ \\
\midrule

% D2 — Data Governance
\multirow{2}{*}{\parbox{1.6cm}{\raggedright D2 Data\newline Governance}}
  & 03 & Data Quality Assurance
  & Ensure training/validation data quality, completeness, representativeness
  & A.7.1--7.3
  & Art.\,10(1--4)
  & \textsc{map}\,2.3;\newline \textsc{me}\,2.6
  & $\square$ & $\blacksquare$ & $\blacksquare$ \\[2pt]

  & 04 & Bias Detection \& Mitigation
  & Monitor for and address unfair bias in data and model outputs
  & A.7.4
  & Art.\,10(2f--g)
  & \textsc{me}\,2.10--2.11
  & $\square$ & $\blacksquare$ & $\blacksquare$ \\
\midrule

% D3 — Human Oversight
\multirow{2}{*}{\parbox{1.6cm}{\raggedright D3 Human\newline Oversight}}
  & 05 & Human Oversight Design
  & Design systems enabling effective human understanding and intervention
  & Cl.\,5;\newline A.15.1--15.2
  & Art.\,14(1--3)
  & \textsc{gv}\,3.2;\newline \textsc{map}\,3.5
  & $\blacksquare$ & $\blacksquare$ & $\square$ \\[2pt]

  & 06 & Override \& Intervention
  & Implement mechanisms for human override and safe system interruption
  & A.15.3--15.4
  & Art.\,14(4--5)
  & \textsc{mg}\,2.4
  & $\blacksquare$ & $\blacksquare$ & $\square$ \\
\midrule

% D4 — Transparency
D4 Transparency
  & 08 & Transparency Communication
  & Provide information on AI system capabilities and limitations
  & A.13--14
  & Art.\,13(2--3)
  & \textsc{gv}\,4.1--4.3;\newline \textsc{map}\,5.2
  & $\blacksquare$ & $\blacksquare$ & $\blacksquare$ \\
\midrule

% D5 — Accuracy & Robustness
\multirow{2}{*}{\parbox{1.6cm}{\raggedright D5 Accuracy \&\newline Robustness}}
  & 09 & Accuracy Validation
  & Validate AI accuracy and performance against defined metrics
  & A.9
  & Art.\,15(1--3)
  & \textsc{me}\,2.5
  & $\blacksquare$ & $\blacksquare$ & $\blacksquare$ \\[2pt]

  & 10 & Robustness \& Security
  & Test resilience to perturbations, adversarial inputs, and cyber threats
  & A.10
  & Art.\,15(4--5)
  & \textsc{me}\,2.6--2.7,\newline 2.9
  & $\blacksquare$ & $\blacksquare$ & $\blacksquare$ \\
\midrule

% D6 — Documentation
D6 Documentation
  & 07 & System Documentation
  & Maintain technical documentation sufficient for audit and compliance
  & Cl.\,7.5;\newline A.12
  & Art.\,11;\newline Annex\,IV
  & \textsc{gv}\,1.3--1.5;\newline \textsc{me}\,2.1
  & $\blacksquare$ & $\blacksquare$ & $\blacksquare$ \\
\midrule

% D7 — Supply Chain
D7 Supply Chain
  & 11 & Supplier AI Assessment
  & Evaluate and monitor third-party AI components and services
  & A.18
  & Art.\,23--25
  & \textsc{gv}\,6.1;\newline \textsc{mg}\,3.1
  & $\square$ & $\square$ & $\blacksquare$ \\
\midrule

% D8 — Incident Management
D8 Incident Mgmt
  & 12 & Incident Response
  & Detect, respond to, and learn from AI system incidents and anomalies
  & Cl.\,10;\newline A.16--17
  & Art.\,9(7), 72
  & \textsc{me}\,3.1;\newline \textsc{mg}\,4.1--4.3
  & $\blacksquare$ & $\blacksquare$ & $\blacksquare$ \\

\bottomrule
\end{tabular}

\vspace{0.4em}
{\scriptsize
$\blacksquare$\,Active \quad $\square$\,Not applicable \qquad
\textbf{Tier key:}
T1\,=\,Vehicle (ADAS, ADS, V2X client);\;
T2\,=\,Edge/RSU (smart signals, pedestrian detection, V2I);\;
T3\,=\,Cloud (TMC, model training, OTA updates).\\[2pt]
\textbf{Abbreviations:}
Cl.\,=\,Clause; A.\,=\,Annex\,A control; Art.\,=\,Article;
\textsc{gv}\,=\,\textsc{govern};
\textsc{map}\,=\,\textsc{map};
\textsc{me}\,=\,\textsc{measure};
\textsc{mg}\,=\,\textsc{manage}.\\[2pt]
\textbf{Note:} UC numbering reflects synthesis order during the crosswalk process rather than sequential domain ordering.}
\end{table*}

\subsection{Worked Example: One Obligation Cluster End to End}
\label{subsec:worked_example}

A methodology stated as rules is hard to check. This subsection follows one obligation cluster through every stage of the pipeline, from the source clauses to the figure the engine prints for a named deployment. Table~\ref{tab:worked} holds the same trace in compressed form, one row per stage. Data quality is the subject for two reasons. It exercises the intent test at the point where that test does the most work, and its evidence is produced by one party while another party answers for it.

\textbf{Extraction.} Twelve atoms enter the cluster. ISO/IEC~42001 contributes four, covering resources for AI systems (A.6.2.3) together with the data management controls A.7.1 to A.7.3. Article~10 of the EU AI Act contributes five, paragraphs~1 to~5. The NIST AI RMF contributes three from the MAP~2 category, subcategories 2.1 to 2.3. Each atom carries its identifier, a one-sentence obligation statement, and the minimal evidence implication needed to demonstrate execution.

\textbf{Coding.} All twelve then receive identical labels, which is the first indication that they belong together. Governance domain D2, tier applicability T2 and T3, and lifecycle phases design, build, and monitor. One of those labels deserves comment. The vehicle tier is excluded during coding and not afterwards, because vehicles consume deployed models and hold none of the training or validation sets whose quality these obligations govern. No vehicle-tier evidence could discharge them.

\textbf{Alignment.} Shared labels are not sufficient for a merge, and the intent test is where the cluster divides. Every atom names the same governance object, the dataset. Not every atom requires the same action on it. Relevance, representativeness, and completeness under Article~10(2) to~10(4) and A.7.1 to A.7.3 are actions on data quality. Examination for unfair bias under Article~10(2f) to~10(2g) and MEASURE~2.11 is an action on model behaviour observed through data. Same object, different action, so the test fails between them and the cluster splits into UC-03 and UC-04. Data governance yields two controls for that reason and not because two seemed convenient.

Within the surviving cluster the strictness test resolves asymmetrically. Article~10 prescribes data governance practice in operational detail, down to examination for biases, identification of gaps, and statements of intended purpose. Annex~A.7 requires data management without prescribing content. Any implementation satisfying Article~10 also satisfies A.7, so Article~10 supplies the control objective while the ISO atoms and the NIST subcategories remain trace-linked subsets. One atom survives outside the merge. Article~10(5) permits processing special categories of personal data for bias detection under stated conditions, and no ISO or NIST atom expresses a comparable permission, so it is retained as a framework-specific extension.

\textbf{Unified control.} What emerges is UC-03, Data Quality Assurance. Its objective covers training and validation data quality, completeness, and representativeness. Activation follows the coding at T2 and T3, and the control carries trace links back to A.7.1--A.7.3, Article~10(1)--10(4), MAP~2.3, and MEASURE~2.6.

\textbf{Evidence.} That control binds to the Data Quality Report, produced at T2 and T3 with T2 as its primary tier. The artifact serves all three frameworks and both controls the cluster produced. Splitting at the alignment stage therefore costs nothing in documentation. Two controls, one artifact.

\textbf{Tool output.} Run against the urban smart intersection, the engine activates UC-03 at T2 under the roadside integrator and at T3 under the transport authority, counts twelve contributing source obligations, and binds the Data Quality Report. Its forward links join the 141 verified in that scenario, none of them broken.

\textbf{Accountability.} The same activation record makes the divergence of Section~\ref{subsec:accountability} concrete instead of hypothetical. Observation locus sits at T2, where roadside sensing generates the data. Legal accountability sits with the transport authority as deployer. Production duty and acceptance duty therefore fall to different organizations, and the evidence exchange contract carrying the Data Quality Report between them is what keeps the obligation discharged and not merely assigned.

\begin{table}[!t]
\centering
\caption{Data quality obligations traced end to end, from source clauses through coding and alignment to the engine output for the urban scenario.}
\label{tab:worked}
\footnotesize
\renewcommand{\arraystretch}{1.15}
\begin{tabular}{@{} L{1.9cm} L{5.6cm} @{}}
\toprule
\textbf{Stage} & \textbf{Result} \\
\midrule
Extraction & 12 atoms: ISO A.6.2.3, A.7.1--A.7.3; EU Art.\,10(1)--10(5); NIST MAP 2.1--2.3 \\
Coding & Domain D2; tiers T2, T3; phases design, build, monitor \\
Intent test & Passes within the cluster; fails against bias atoms, which split to UC-04 \\
Strictness test & Asymmetric. Article~10 dominates; ISO and NIST atoms trace-linked \\
Extension & Art.\,10(5) retained; no ISO or NIST counterpart \\
Unified control & UC-03 Data Quality Assurance, active at T2 and T3 \\
Evidence & Data Quality Report; primary tier T2; serves 3 frameworks, 2 controls \\
Engine output & Activated T2 (integrator), T3 (authority); 12 source obligations; 0 broken links \\
Accountability & Observation locus T2, legal accountability T3; evidence exchange contract required \\
\bottomrule
\end{tabular}
\end{table}

\section{ITS Component Mapping}
\label{sec:mapping}

ITS deployments distribute AI behavior and evidence across vehicle, roadside, and cloud tiers, often under different operational owners. Governance fails when controls are assigned to the wrong tier or when evidence expectations assume data access that cannot be negotiated in practice. This section maps UGAF-ITS unified controls (Table~\ref{tab:unified_controls}) to representative ITS AI components and identifies the evidence sources that make each control auditable where behavior executes. The mapping is representative and not exhaustive, and it is intended to guide responsibility allocation under multi-stakeholder delivery models. Table~\ref{tab:uc_component_map} summarizes the component-to-control mapping and evidence sources by tier.

\begin{table*}[!t]
\centering
\caption{ITS deployment tiers, unified controls, and governance evidence sources.
Primary UCs are listed first; trace dependencies noted with $\hookrightarrow$.
Evidence sources are ordered from operational logs to strategic compliance artifacts.}
\label{tab:uc_component_map}
\footnotesize
\setlength{\tabcolsep}{4pt}
\renewcommand{\arraystretch}{1.35}
\begin{tabular}{@{} c L{3.6cm} L{2.6cm} L{5.2cm} L{2.4cm} @{}}
\toprule
\textbf{Tier} &
\textbf{Components} \newline {\scriptsize (examples)} &
\textbf{Unified Controls} &
\textbf{Evidence Sources} \newline {\scriptsize (examples, operational $\to$ strategic)} &
\textbf{Owner} \\
\midrule

\textbf{T1} \newline Vehicle &
ADAS perception, driver monitoring, V2X client, local planning under ODD constraints &
\textbf{UC-05, UC-06,} \newline \textbf{UC-09, UC-10} \newline {\scriptsize $\hookrightarrow$ UC-07} &
Oversight Protocol; Intervention \& Fallback Plan; Event \& Override Logs; V\&V Test Report (ODD-scoped); Security Test Report; Release Manifest &
\textbf{OEM} \newline {\scriptsize vehicle operator} \\
\midrule

\textbf{T2} Edge/RSU &
VRU detection, adaptive signal logic, V2I gateways (e.g., SPaT/MAP), local decision engines &
\textbf{UC-03, UC-04,} \newline \textbf{UC-08, UC-12} \newline {\scriptsize $\hookrightarrow$ UC-06, UC-10} &
Data Quality Report; Bias \& Performance Report; Configuration Baseline (site + calibration); RSU Decision Logs; Monitoring \& Alerting Plan; Escalation Record &
\textbf{Road authority} \newline {\scriptsize + integrator} \\
\midrule

\textbf{T3} \newline Cloud &
TMC analytics, cross-site monitoring, model registry, training/evaluation pipelines, OTA distribution &
\textbf{UC-01, UC-02,} \newline \textbf{UC-07, UC-11,} \newline \textbf{UC-12} \newline {\scriptsize $\hookrightarrow$ UC-09, UC-10} &
Risk Register + Treatment Log; Technical File; Supplier Audit Report; Third-Party Component Register; Incident Ticket Log; Model Change Notice; Deployment Record; Post-Deployment Monitoring Report &
\textbf{Authority ops} \newline {\scriptsize / cloud operator} \\

\bottomrule
\end{tabular}
\end{table*}

\subsection{Tier-Level Control Allocation}
\label{subsec:tier_controls}

Section~\ref{subsec:its_layer} defined the governance rationale for each tier. This subsection specifies which controls attach to which components and what evidence makes them auditable.

\textbf{Vehicle tier.} Controls UC-05 (Human Oversight Design) and UC-06 (Override \& Intervention) anchor the human interaction contract through clear status cues, safe interruption mechanisms, and operator-facing procedures. UC-09 (Accuracy Validation) and UC-10 (Robustness \& Security) bind performance and resilience to the declared Operational Design Domain and to adversarial or malformed inputs that can arise through sensors or communications. The practical boundary is simple: if a control cannot be executed deterministically on-board, it cannot be relied upon for real-time safety claims. UC-07 (System Documentation) is satisfied by trace links from vehicle-tier records into the centralized Technical File maintained at the cloud tier.

\textbf{Edge/RSU tier.} UC-03 (Data Quality Assurance) and UC-04 (Bias Detection \& Mitigation) attach directly to roadside sensing and aggregation, where calibration, occlusion, illumination, and sensor health determine whether downstream outputs remain valid across time of day and traffic mix. UC-08 (Transparency Communication) governs how intersection guidance and limitations are communicated to downstream actors, while UC-12 (Incident Response) begins operationally at the edge through anomaly triggers and escalation discipline. This tier is a governance hotspot because it sits at an organizational boundary, and because failures often manifest as externally visible behavior even when root causes originate in sensing quality, configuration drift, or model degradation.

\textbf{Cloud tier.} UC-01 (AI Risk Assessment) and UC-02 (Risk Treatment) are anchored here because credible risk decisions require aggregated signals, consistent acceptance criteria, and traceable mitigation actions across intersections and versions. UC-07 (System Documentation) is operationalized through the Technical File, which links requirements, design decisions, datasets, evaluation evidence, deployment records, and change notices into an assessor-navigable record. UC-11 (Supplier AI Assessment) tracks third-party model provenance, contractual constraints, and downstream obligations~\cite{javeed2025malware}. UC-11 is scoped to T3 because supplier assessment is an organizational-level activity. Vendor selection, contractual AI governance clauses, audit scheduling, and component acceptance happen at the procuring entity (typically the cloud operator or transport authority), not at the edge or in the vehicle. The assessed components may deploy to any tier, but the assessment decision sits at T3. Its outcome propagates downward through the Technical File (UC-07) and deployment validation (UC-09). This keeps supplier governance evidence reachable from all tiers without forcing each tier operator to run independent supplier audits.

\subsection{Accountability Under Multi-Party Ownership}
\label{subsec:accountability}

Tier placement answers where a control can be enforced and where its evidence becomes observable. Who must answer for that evidence is a separate question, and the architecture does not settle it. Governance models built on architecture tend to fail at this seam, because a compute boundary is an engineering fact while accountability is a contractual and legal one.

UGAF-ITS therefore separates four attributes that a coarse reading of the tier model collapses into one, and the separation is what keeps the seam visible. Execution locus names the tier where AI behavior runs. Observation locus names the tier where supporting evidence is generated and retained. Operational control names the party able to change system behavior. Legal accountability names the party bearing the regulatory duty. Tier allocation fixes the first two, both of which follow from the architecture. The remaining two are declared per deployment, and Table~\ref{tab:accountability} records how the framework responds when they diverge from the tier assignment.

None of this is a modelling convenience, because the distinction carries legal weight. Provider duties under Article~16 and deployer duties under Article~26 attach to roles, not to hardware. Under Article~25, a distributor, importer, deployer, or other third party becomes a provider once it puts its name on a high-risk system already placed on the market, makes a substantial modification, or changes the intended purpose so that the system becomes high-risk~\cite{euaiact}. A roadside integrator that rebrands a vendor perception stack crosses that line. So does a transport authority that retrains a supplied model on local traffic. Compute placement is unchanged in both cases. The accountable party is not.

Three configurations recur in ITS delivery. Each divides a different pair of the four attributes, and each therefore resolves differently.

\textbf{Shared responsibility.} Evidence for one control is produced at a tier operated by a party that does not hold the duty. Data quality assurance (UC-03) is the standard ITS case, since roadside sensing generates the data, a cloud operator trains on it, and a road authority owns the roadside unit. The framework splits such a control into a production duty at the observation locus and an acceptance duty at the accountable party, bound by an evidence exchange contract. Both endpoints are recorded, so a production duty with no matching acceptance duty surfaces as ownership ambiguity (F4) instead of resolving silently to whichever operator runs the hardware.

\textbf{Outsourced components.} Contracting changes the picture again. A managed-service vendor may operate an edge component while the road authority remains the deployer. Operational control transfers. Legal accountability does not transfer by default, and Article~25 can move it in the opposite direction from the contract when the vendor rebrands or substantially modifies what it operates. UC-11 (Supplier AI Assessment) is the mechanism here, and its allocation to the cloud tier follows from this reasoning and not from where the assessed component executes. Vendor selection, contractual governance clauses, audit scheduling, and component acceptance are organizational activities performed by the procuring entity, while the assessed component may run at any tier.

\textbf{Cross-tier ownership.} The third configuration removes the single owner altogether. One AI function can span tiers, as when a model is trained in the cloud, deployed to the edge, and consumed in the vehicle. The framework instantiates the control at every tier where obligation-relevant behavior occurs and designates one instance as lead for evidence consolidation. Shared trace identifiers keep the instances joined, and the chains in Section~\ref{subsec:cross_tier} carry them across the boundaries.

A fourth configuration resists resolution inside the model altogether. An accountable party may have no tier presence at all, as with a model supplier that ships weights and operates nothing, and its obligations then generate evidence that can arrive only through contractual flow-down. UGAF-ITS reports this as ownership ambiguity instead of assigning the duty to whichever tier happens to host the resulting artifact. Detecting the ambiguity falls within scope. Resolving it requires contract instruments the framework does not supply.

The claim the tier model supports is therefore narrower than architectural determinism. Tier allocation is necessary for defensible evidence, because a duty assigned where the behavior cannot be observed yields no record an auditor can use. Tier allocation is not sufficient for accountability, which requires a declared role assignment carried alongside the architecture in the deployment descriptor.

\begin{table}[!t]
\centering
\caption{Divergence between architectural tier and accountability, and how UGAF-ITS responds. Attributes: E = execution locus, O = observation locus, C = operational control, A = legal accountability.}
\label{tab:accountability}
\footnotesize
\renewcommand{\arraystretch}{1.15}
\begin{tabular}{@{} L{1.75cm} c L{2.45cm} L{2.05cm} @{}}
\toprule
\textbf{Configuration} & \textbf{Diverges} & \textbf{Framework response} & \textbf{Declared outside the model} \\
\midrule
Shared responsibility & A $\neq$ O & Split into production duty at O and acceptance duty at A & Evidence exchange contract \\
Outsourced operation & C $\neq$ A & UC-11 assessment at the procuring entity; Article~25 reclassification checked & Service contract and rebranding terms \\
Cross-tier ownership & E spans tiers & Control instantiated per tier with one lead instance & Consolidation lead and retention terms \\
Absent-tier accountability & A has no tier & Reported as ownership ambiguity (F4) & Contractual evidence flow-down \\
\bottomrule
\end{tabular}
\end{table}

\subsection{Cross-Tier Coordination}
\label{subsec:cross_tier}

Cross-tier governance breaks in two predictable ways. Evidence can remain trapped inside tier-specific tools, preventing central teams from justifying decisions under audit. Alternatively, evidence can be centralized in ways that strip context, so logs no longer reflect the configuration and sensing conditions that produced them. Two invariants are therefore required: every observation stays bound to its tier, site, and versioned baseline; and every escalation carries enough metadata to support reconstruction without requiring raw-log pooling.

Update governance introduces the second coupling. Cloud releases propagate to edge and then to vehicles, while accountability must follow the same path. Documentation, validation evidence, and security checks must reference the deployed configuration, not a laboratory proxy.

UC-12 operationalizes this coordination without assuming vertical integration. A roadside trigger fires when an RSU detects a sustained rise in vulnerable road user false positives, a sensor-health drift alert, or inconsistent intersection-state messaging. The edge tier records the event in RSU Decision Logs, binds it to the current Configuration Baseline, and attaches relevant indicators from the Data Quality Report. Escalation sends a signed incident summary plus artifact hashes, so the cloud tier can verify integrity while leaving raw traces under local retention policies. Cloud operations open an entry in the Incident Ticket Log, correlate site-level reports, classify severity against the Incident Response Plan, and select mitigation as a controlled change. The decision updates the Risk Register and Risk Treatment Log, and a Model Change Notice captures the rationale, the supporting evidence, the affected tiers, and the rollout and rollback conditions. Closure requires confirmation in a Post-Deployment Monitoring Report that the anomaly no longer manifests under operational conditions, plus updates to thresholds, test cases, or playbooks that prevent recurrence. This chain keeps UC-07 documentation integrity synchronized with UC-01/UC-02 risk governance while respecting multi-vendor evidence constraints.

\section{Evaluation}\label{sec:validation}
 
Claims about a governance framework are only as strong as the procedure that tests them. This section states the evaluation design before reporting any result, so that each number carries an explicit account of what it can and cannot support. An open-source tool, the UGAF-ITS Governance Engine, encodes the crosswalk catalog, the evidence backbone, and the tier-allocation logic as executable computations. Four architecturally distinct ITS deployment scenarios then exercise those computations, and Section~\ref{subsec:results} reports results for coverage, consolidation, evidence reuse, traceability, and cross-tier dependency analysis.
 
One behaviour is worth stating before the design. Structural properties of the catalog hold stable across full three-tier deployments and contract predictably when tiers are removed. When a deployment lacks a tier, the controls requiring that tier's evidence sources become inapplicable, and the engine reports which obligations are affected and why. That contraction is the intended behaviour of an allocation model and not a defect, though it is also the reason coverage figures cannot be read as compliance figures.

\subsection{Evaluation Design and Levels of Evidence}
\label{subsec:eval_design}

Evidence about a governance framework arrives at three distinct levels. Conflating them is the most common route to overstating what such work has shown. Table~\ref{tab:evidence_levels} separates the three. Levels~L1 and~L2 are performed here. Level~L3 is not.

Internal verification (L1) asks whether the encoded catalog is self-consistent. The engine traverses every obligation-to-control and control-to-artifact link and reports orphans or breaks. A clean result confirms the encoding. It carries no information about whether the encoded mappings match expert judgment.

Scenario demonstration (L2) asks whether structural properties of the catalog hold when deployment architecture varies. Four scenarios spanning three-tier to single-tier topologies exercise the same computations. A stable result confirms designed behaviour under architectural variation. It does not establish that an organization adopting the framework would satisfy an auditor.

External validation (L3) would require independent expert assessment of crosswalk correctness, practitioner audits conducted with the engine, or deployment inside an operational ITS programme. Section~\ref{subsec:threats} sets out what remains outstanding at this level and what would close it.

Three terms carry the distinction through the remainder of the paper. \emph{Scoped framework coverage} names the fraction of extracted obligations addressed by activated controls within the crosswalk's modelling boundary. \emph{Structural validation} names L1 and L2 results taken together. \emph{Prototype governance support} names what the engine offers a practitioner today. None of the three is a claim about operational compliance readiness, which depends on implementation quality, organizational maturity, and audit judgment.

\begin{table*}[!t]
\centering
\caption{Levels of evidence for a governance framework. Levels~L1 and~L2 are performed in this paper; level~L3 is not.}
\label{tab:evidence_levels}
\footnotesize
\renewcommand{\arraystretch}{1.15}
\begin{tabular}{@{} l L{3.1cm} L{4.4cm} L{4.6cm} c @{}}
\toprule
\textbf{Level} & \textbf{Question answered} & \textbf{Procedure} & \textbf{Cannot support} & \textbf{Status} \\
\midrule
L1 Internal verification & Is the encoded catalog self-consistent? & Exhaustive traversal of forward and reverse trace links across 154 obligations, 12 controls, and 20 artifacts & Correctness of the mappings against expert judgment & Performed \\
L2 Scenario demonstration & Do structural properties hold as architecture varies? & Identical computations executed across four deployment archetypes from three-tier to single-tier & Audit adequacy or usability in a live deployment & Performed \\
L3 External validation & Do practitioners agree, and does the framework work in the field? & Independent expert assessment, practitioner audits with the engine, operational deployment & Not applicable, since the level is not performed & Outstanding \\
\bottomrule
\end{tabular}
\end{table*}
 
\subsection{Governance Engine Architecture}\label{subsec:tool_arch}
 
The UGAF-ITS Governance Engine is a Python package that takes a deployment descriptor as input and produces a compliance report as output. The deployment descriptor is a YAML file specifying the system's name, its AI components, and for each component: the tier it occupies (T1~vehicle, T2~edge/RSU, T3~cloud), its risk level, and the responsible owner. This representation is minimal by design. It captures the architectural decisions that determine where governance attaches without requiring operational data, proprietary logs, or access to running systems. A minimal descriptor takes the following form:

\begin{verbatim}
system_name: "Rural Intersection"
components:
  - name: "Pedestrian Detection"
    tier: T2_EDGE
    risk_level: high
    owner: "Road Authority"
  - name: "Signal Controller AI"
    tier: T2_EDGE
    risk_level: high
    owner: "Road Authority"
\end{verbatim}
 
The engine executes eight computations against the knowledge base, which encodes the complete UGAF-ITS catalog: 154 source obligations decomposed from ISO/IEC~42001, the EU AI Act, and the NIST AI RMF; 12 unified controls with clause-level traceability; 20 evidence artifacts with multi-framework reuse properties; and the tier-activation matrix from Table~\ref{tab:unified_controls}.
 
\begin{enumerate}[label=\textbf{C\arabic*.},leftmargin=2em,itemsep=2pt]
\item \textbf{Tier-aware control activation} determines which unified controls are activated based on which tiers have components. A control activates only when at least one of its designated tiers is populated. This is the computable form of the principle that governance attaches where behavior executes.
 
\item \textbf{Evidence backbone with cross-framework reuse} computes the minimal artifact set required for the activated controls and compares it against the enumerated siloed register of \ref{app:register}. Both sides obey one rule. A document or artifact counts when it is producible at an active tier and evidences at least one activated control. A three-tier deployment is therefore compared against all 37 documents, a two-tier deployment against 36, and a single-tier deployment against 16.
 
\item \textbf{Bidirectional traceability verification} traverses every forward link (source obligation $\rightarrow$ unified control $\rightarrow$ evidence artifact) and every reverse link (artifact $\rightarrow$ control $\rightarrow$ obligations) to confirm that no mapping is broken. A single broken link produces a named diagnostic.
 
\item \textbf{Per-framework coverage analysis} computes the fraction of each framework's scoped obligations that are covered by the activated control set. Gaps are classified by cause instead of being treated as undifferentiated failures.
 
\item \textbf{Gap detection and classification} categorizes every uncovered obligation into one of four classes: organizational procedure (internal audit and management review workflows that operate outside technical controls), regulatory workflow (conformity assessment, registration, and marking obligations that execute outside the engineering scope), context-setting governance (organizational profiling and stakeholder engagement that precede control instantiation), and tier-not-present (obligations targeting a tier absent from the deployment).
 
\item \textbf{Crosswalk consolidation analysis} quantifies how the 154 source obligations consolidate into unified controls, broken down by governance domain and by contributing framework. This computation answers the question a reviewer will ask: is the 154-to-12 consolidation a trivial relabelling, or does it involve genuine analytical synthesis across heterogeneous sources?
 
\item \textbf{Cross-tier dependency analysis} maps evidence chain dependencies between tiers: sequences of controls and artifacts that span organizational boundaries. For example, the incident response escalation chain traverses edge anomaly detection (UC-12), cloud triage and risk update (UC-01, UC-02), controlled model change (UC-07), and redeployment validation (UC-09). These chains are the operational structures that siloed compliance approaches cannot represent.
 
\item \textbf{Structured comparison} produces the M1--M5 metrics (coverage, duplication reduction, evidence reuse, traceability, evidence-package structure) in a format suitable for direct paper inclusion.
\end{enumerate}
 
\begin{figure*}[!t]
\centering
\includegraphics[width=\textwidth]{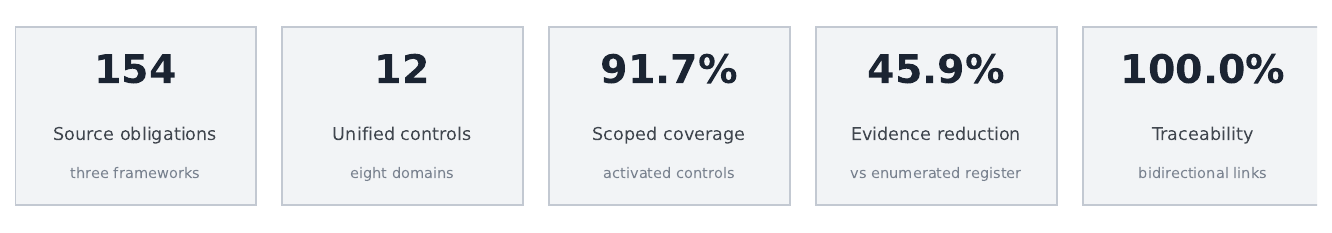}
\caption{Headline metrics for the urban smart intersection scenario, also presented in the engine's HTML dashboard. The panel is generated from the JSON reports by \texttt{python make\_figures.py} after \texttt{python run\_validation.py}, so every value shown is read from the run and not transcribed. Source code, scenarios, the siloed baseline register, and the dashboard are available at \url{https://doi.org/10.5281/zenodo.19339071}.}
\label{fig:tool_dashboard}
\end{figure*}

Four output formats leave the engine. JSON reports carry the full computation for programmatic consumption, Markdown summaries present the same content in a form an auditor can read, a comparative table aggregates metrics across scenarios, and an HTML dashboard presents the whole result set for inspection. Every figure reporting results in this paper is generated from the JSON reports by a separate script, so a plotted value cannot drift from a reported one. Figure~\ref{fig:tool_dashboard} shows the headline metrics for the primary urban scenario.

The dashboard is a single self-contained file. Charts are inline SVG generated from the same reports, interaction runs on data embedded in the page, and no external script, font, or network access is required. A reader who downloads the archive can open it directly and navigate the control catalog, the evidence backbone, the siloed register, the granularity analysis, and the classified gaps without installing anything. A scenario builder in the same page composes a deployment descriptor and emits the YAML.

Extension beyond the four scenarios reported here requires only that descriptor, naming components, tiers, risk levels, and owners. No operational data, proprietary logs, or access to running systems is needed.
 
The tool is open-source under the MIT license and requires Python~3.8 or later with PyYAML, plus Matplotlib for figure generation. Scenario results, the dashboard, and the reported metrics reproduce with a single command. A Zenodo DOI accompanies the repository for long-term archival.

\subsection{Deployment Scenarios}\label{subsec:scenarios}
 
Governance under distribution is sensitive to architectural shape. A framework exercised against a single topology cannot claim generality. We evaluate UGAF-ITS against four ITS deployment scenarios that vary systematically along the dimensions that most affect governance: the number of active tiers, the number of AI components, the number of independent stakeholders, and the risk profile of the deployed components.
 
How those scenarios were built determines what any result drawn from them can mean, so the construction is worth stating before the descriptions. All four are synthetic architectural descriptors written by the authors. Component inventories, tier assignments, ownership splits, and risk classifications follow publicly documented ITS deployment patterns and the tier structures defined in ISO~21217 and ARC-IT~\cite{iso21217,arcit}. No operator supplied deployment data, and no operator reviewed the resulting descriptors. The scenarios are therefore semi-realistic in structure and fully synthetic in content.

That construction fixes what the results can generalize over. The four descriptors sample an architectural space defined by tier count, component density, stakeholder count, and risk profile, so the results characterize how the framework behaves across that space. They do not sample a population of real deployments, and scenario realism itself was never tested. A framework property that holds across all four scenarios is a property of the framework under architectural variation and not evidence about how frequently such architectures occur or how a real operator would populate one. Section~\ref{subsec:threats} treats the consequences for external validity.

\textbf{Scenario~1: Urban smart intersection.} A high-maturity metropolitan intersection whose stakeholder structure follows the multi-operator pattern documented in Dubai RTA's published ITS strategy~\cite{RTA_AIStrategy2030}. Twelve AI components span all three tiers under three independent owners (OEM, roadside integrator, transport authority). All components are high-risk. This scenario represents maximum governance complexity: full tier coverage, maximum stakeholder fragmentation, and the highest evidence coordination burden.
 
\textbf{Scenario~2: Highway corridor ADS.} The second scenario keeps the tier structure and thins it out. Advanced driving support systems sit at the vehicle tier, roadside monitoring at the edge, and centralized fleet analytics at the cloud tier. Seven components span all three tiers under three owners. This scenario preserves the three-tier topology of Scenario~1 but with fewer components per tier, testing whether governance properties hold when tier populations are sparser.
 
\textbf{Scenario~3: Transit priority corridor.} Removing a tier is the next variation. A bus rapid transit corridor carries edge-tier signal priority logic and cloud-tier scheduling optimization. No vehicle-tier AI components are present; the transit vehicles are treated as passive consumers of intersection state. Five components span two tiers under two owners. This scenario tests the framework's behavior when an entire tier is absent and the governance footprint contracts accordingly.
 
\textbf{Scenario~4: Rural intersection.} The last scenario strips the architecture to its minimum. Two edge-tier components sit under a single owner, a basic pedestrian detection unit and a simple signal controller. No vehicle or cloud tier. This is the stress test: minimum architecture, minimum stakeholders, and the sharpest question to the framework: does it degrade honestly instead of producing meaninglessly high scores for a deployment that barely qualifies as distributed?

\subsection{Results}\label{subsec:results}

Every figure below sits at level~L1 or level~L2 of Table~\ref{tab:evidence_levels}. Traceability results report internal verification. Coverage, reduction, reuse, and dependency results report scenario demonstration. None of them reports operational compliance readiness, and the wording throughout keeps to the scoped terms fixed in Section~\ref{subsec:eval_design}.

Table~\ref{tab:comparative_validation} presents the comparative results across all four scenarios. We report metrics across five evaluation dimensions: framework coverage (M1), duplication reduction (M2), evidence reuse (M3), traceability (M4), and instantiation depth (a new metric introduced in this evaluation to distinguish structural coverage from architectural completeness).
 
\begin{table}[!t]
\centering
\caption{Multi-scenario comparative evaluation results. Metrics span five
evaluation dimensions across four ITS deployment scenarios. The siloed
baseline counts documents from the register of \ref{app:register} that are
producible at each scenario's active tiers.}
\label{tab:comparative_validation}
\footnotesize
\renewcommand{\arraystretch}{1.12}
\setlength{\tabcolsep}{3.5pt}
\resizebox{\columnwidth}{!}{%
\begin{tabular}{@{} l l c c c c @{}}
\toprule
\textbf{Dimension} & \textbf{Metric} &
\textbf{Urban} & \textbf{Highway} & \textbf{Transit} & \textbf{Rural} \\
\midrule

\multirow{3}{*}{\textit{Architecture}}
  & Active tiers          & T1--T3 & T1--T3 & T2, T3 & T2 \\
  & AI components         & 12 & 7 & 5 & 2 \\
  & Stakeholders          & 3  & 3 & 2 & 1 \\
\midrule

\multirow{4}{*}{\textit{M1: Coverage}}
  & ISO/IEC 42001         & 94.5\% & 94.5\% & 94.5\% & 92.7\% \\
  & EU AI Act             & 91.9\% & 91.9\% & 91.9\% & 89.2\% \\
  & NIST AI RMF           & 88.7\% & 88.7\% & 88.7\% & 82.3\% \\
  & \textit{Average}      & 91.7\% & 91.7\% & 91.7\% & 88.1\% \\
\midrule

\multirow{3}{*}{\textit{M2: Reduction}}
  & Unified artifacts     & 20     & 20     & 18     & 12 \\
  & Siloed baseline       & 37     & 37     & 36     & 16 \\
  & Reduction             & 45.9\% & 45.9\% & 50.0\% & 25.0\% \\
\midrule

\multirow{2}{*}{\textit{M3: Reuse}}
  & Avg.\ fw/artifact     & 2.80   & 2.80   & 2.89   & 2.92 \\
  & 3-fw artifacts        & 80.0\% & 80.0\% & 88.9\% & 91.7\% \\
\midrule

\textit{M4: Traceability}
  & Bidirectional         & 100\%  & 100\%  & 100\%  & 100\% \\
\midrule

\multirow{3}{*}{\textit{Depth}}
  & Instantiation         & 0.94 & 0.68 & 0.49 & 0.46 \\
  & Cross-tier chains     & 5    & 5    & 4    & 0 \\
  & Interpretation        & Prod. & Subst. & Part. & Part. \\
\bottomrule
\end{tabular}%
}
\end{table}
 
\subsubsection{Framework Coverage (M1)}\label{subsec:coverage}
 
Scoped framework coverage remains stable across all three-tier deployments at 91.7\% average: 94.5\% for ISO/IEC~42001, 91.9\% for the EU AI Act, and 88.7\% for the NIST AI RMF. Rural drops to 88.1\% because the absence of vehicle and cloud tiers deactivates UC-11 (Supplier AI Assessment), which requires cloud-tier infrastructure. The coverage gaps that persist across all scenarios fall into three categories that resist technical harmonization. ISO gaps arise from internal audit and management review workflows. EU AI Act gaps arise from conformity assessment interactions, registration, and marking obligations. NIST gaps arise from organizational profiling and stakeholder engagement activities that precede control instantiation. These are not framework weaknesses. They are the boundary where technical governance ends and organizational procedure begins.

The informative content of M1 lies in the 8\% gap, not the 92\% coverage. The gap classifies precisely which obligation types resist technical harmonization and why.

Two qualifications bound the metric. M1 reports \emph{scoped framework coverage} within the crosswalk's modeling boundary: the fraction of extracted obligations addressed by activated controls. It does not measure operational compliance readiness, which depends on implementation quality, organizational maturity, and audit judgment. M1 also treats all mapped obligations equally. A mandatory EU AI Act high-risk obligation carries the same weight as a voluntary NIST subcategory. A weighted coverage metric, one that reflects legal bindingness, risk severity, or implementation complexity, could provide a more operationally meaningful signal. Building such a metric requires stakeholder-validated weight assignments that fall outside the current scope.
 
\subsubsection{Evidence Reduction (M2)}
 
UGAF-ITS reduces the evidence pack from 37 siloed documents to 20 unified artifacts in three-tier deployments, a 45.9\% reduction. Any such figure is only as credible as the baseline it is measured against, so that baseline is enumerated here instead of estimated. \ref{app:register} lists all 37 documents with the framework that produces each one, the clause or article that requires it, the tiers at which it is generated, and the unified controls it would evidence. Fifteen documents come from ISO/IEC~42001 documented-information requirements across Clauses~4--10 and the Annex~A control areas, twelve from EU AI Act Annex~IV together with the record-keeping duties of Articles~9 to~17 and Articles~47, 49, and~72, and ten from artifact types named across the four NIST AI RMF functions. The governance engine computes the baseline from that register at run time, so any reader can change a document attribution and observe the effect on every reported reduction figure.

Both sides of the comparison obey one rule. A document or artifact counts for a deployment when it is producible at an active tier and it evidences at least one activated control. Applying different rules to the unified and siloed sides would let the comparison be tuned, which is the objection an enumerated baseline exists to answer.

Scoping changes the comparison less than architectural intuition suggests. Transit compares against 36~documents and not~37, because only one document in the register is produced exclusively at the vehicle tier, the instructions for use required by Article~13. Rural compares against~16. Reduction therefore ranges from 25.0\% to 50.0\% across the four archetypes, and it does not increase monotonically with tier count. Transit consolidates marginally more than the three-tier deployments, at 50.0\% against 45.9\%.

That ordering is informative and not anomalous. Siloed documentation is dominated by organizational records that a cloud tier produces whatever sits below it, including policies, scope statements, audit programmes, management reviews, and conformity paperwork. Removing the vehicle tier removes several unified artifacts while removing almost nothing from the siloed pack. A small deployment does not escape the paperwork, which is the condition consolidation is meant to relieve. The result to report is that consolidation holds across the architectural space, between a quarter and a half of the documentation set, rather than that it rewards complexity.

Sensitivity remains worth stating. Under a conservative register of 33~documents the three-tier reduction drops to 39.4\%, and under a generous register of 41 it rises to 51.2\%. Every reduction figure reported here is a function of the register, and the register is now open to inspection and revision.
 
\subsubsection{Evidence Reuse (M3)}
 
Each artifact serves an average of 2.8 frameworks in three-tier deployments. Eighty percent of artifacts in the urban and highway scenarios serve all three frameworks simultaneously. The reuse factor increases slightly as deployments shrink (2.92 for rural) because the remaining artifacts are the cross-cutting ones (risk register, technical file, incident response plan) that inherently serve all three sources. Reuse is bounded by audit constraints: artifacts must preserve tier context, configuration binding, and version metadata. The evidence backbone enforces these constraints by design.
 
\subsubsection{Traceability (M4)}
 
Bidirectional traceability scores 100\% across all four scenarios: zero broken forward links (obligation $\rightarrow$ control $\rightarrow$ artifact) and zero broken reverse links. This result is structural and not coincidental. The knowledge base encodes every obligation-to-control mapping and every control-to-artifact assignment as explicit data, and the traceability computation traverses every link exhaustively. A broken link would indicate a knowledge base defect, not a framework limitation.
 
\subsubsection{Instantiation Depth}
 
Structural coverage tells a reviewer whether obligations map to controls. It does not tell whether a deployment can actually produce the evidence those controls demand. Instantiation depth addresses that question. It is a weighted composite of three normalized factors:

\begin{equation}
D = w_d \cdot d_c + w_r \cdot r_h + w_s \cdot s_b
\label{eq:depth}
\end{equation}

\noindent $d_c$ is component density: the average number of AI components per active tier, normalized by a reference density of 4 components per tier and capped at 1.0. $r_h$ is risk coverage: the fraction of components classified as high-risk. $s_b$ is stakeholder breadth: the number of independent owners, normalized by the maximum observed across scenarios. The default weights are $w_d = 0.50$, $w_r = 0.20$, $w_s = 0.30$. Component density carries the most weight because it determines how much evidence each tier can actually produce. Stakeholder coordination and risk profile contribute proportionally. All weights and normalization parameters are configurable in the governance engine source.

Urban scores 0.94 (production-grade). Highway scores 0.68 (substantial). Transit and rural score below 0.50 (partial). The gradient is informative: two deployments with identical structural coverage (91.7\% for urban and highway) differ meaningfully in their ability to instantiate governance, because highway has fewer components per tier and therefore thinner evidence sources. Depth and structural coverage together provide a two-dimensional characterization that neither metric offers alone.

Weight selection affects absolute scores but not the ranking. We tested $w_d \in [0.30, 0.70]$, $w_r \in [0.10, 0.30]$, and $w_s \in [0.10, 0.40]$ (constrained to $w_d + w_r + w_s = 1$). Under every configuration, the ordering holds: Urban $>$ Highway $>$ Transit $\geq$ Rural. The numbers move. The ranking does not.
 
\begin{figure*}[!t]
\centering
\includegraphics[width=\textwidth]{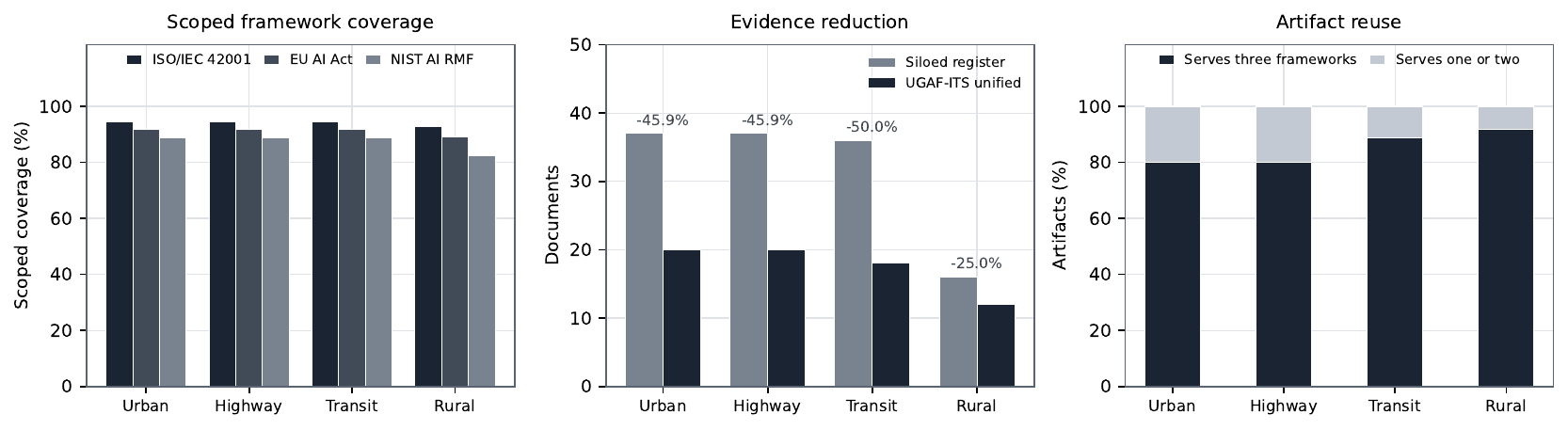}
\caption{Top: framework coverage across four ITS deployment archetypes; bars represent ISO/IEC~42001, EU AI Act, and NIST AI RMF coverage per scenario. Left bottom: evidence document reduction from siloed baseline (scope-aware) to UGAF-ITS unified artifacts per scenario. Right bottom: artifact reuse distribution showing 80\% of artifacts serve all three frameworks simultaneously in three-tier deployments.}
\label{fig:coverage_reduction}
\end{figure*}

Figure~\ref{fig:coverage_reduction} presents the framework coverage distribution across scenarios alongside the evidence reduction and artifact reuse analysis.
 
\subsubsection{Crosswalk Consolidation}
 
The consolidation analysis reveals that the 154-to-12 consolidation is not uniformly distributed. Table~\ref{tab:consolidation_domain} breaks down the contribution per governance domain. Risk Management (D1) consolidates 53 source obligations into 2 unified controls, a 96\% reduction, because all three frameworks converge on risk identification, evaluation, and treatment. The overlap is deep and semantically consistent. Documentation (D6) consolidates 16 obligations into 1 control (94\%) because technical documentation requirements, while phrased differently across frameworks, demand the same underlying evidence structure. Human Oversight (D3) consolidates least aggressively among the domains holding more than one control, at 87\%, because the EU AI Act specifies intervention mechanisms at a level of detail that ISO/IEC~42001 and the NIST AI RMF do not. The unified controls preserve those framework-specific extensions rather than absorbing them.
 
\begin{figure*}[!t]
\centering
\includegraphics[width=\textwidth]{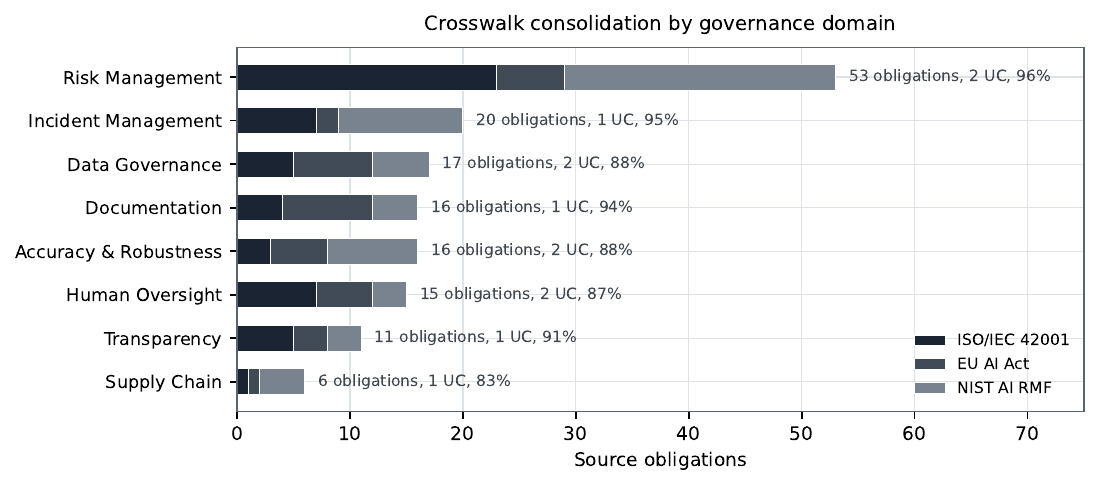}
\caption{Crosswalk consolidation by governance domain. Stacked bars show the contribution of each source framework (ISO/IEC~42001, EU AI Act, NIST AI RMF) to the source obligation count per domain. All eight domains exhibit tri-framework convergence, confirming that unified controls synthesize genuine cross-framework overlap rather than relabelling single-framework requirements.}
\label{fig:consolidation_domain}
\end{figure*}

Every domain exhibits tri-framework alignment. No unified control draws from fewer than all three source frameworks. This confirms that the crosswalk methodology produces genuine harmonization and not framework-specific relabelling. Figure~\ref{fig:consolidation_domain} visualizes the per-domain consolidation with framework-stacked obligation counts.
 
\begin{table}[!t]
\centering
\caption{Crosswalk consolidation by governance domain. All domains exhibit tri-framework alignment. Consolidation ratio = (1 $-$ UCs / source obligations) $\times$ 100.}
\label{tab:consolidation_domain}
\footnotesize
\renewcommand{\arraystretch}{1.15}
\setlength{\tabcolsep}{3.5pt}
\begin{tabular}{@{}l r r r r r r@{}}
\toprule
\textbf{Domain} & \textbf{ISO} & \textbf{EU} & \textbf{NIST} & \textbf{Total} & \textbf{UCs} & \textbf{Ratio} \\
\midrule
D1 Risk Management       & 23 &  6 & 24 & 53 & 2 & 96\% \\
D2 Data Governance        &  5 &  7 &  5 & 17 & 2 & 88\% \\
D3 Human Oversight        &  7 &  5 &  3 & 15 & 2 & 87\% \\
D4 Transparency           &  5 &  3 &  3 & 11 & 1 & 91\% \\
D5 Accuracy \& Robustness &  3 &  5 &  8 & 16 & 2 & 88\% \\
D6 Documentation          &  4 &  8 &  4 & 16 & 1 & 94\% \\
D7 Supply Chain           &  1 &  1 &  4 &  6 & 1 & 83\% \\
D8 Incident Management    &  7 &  2 & 11 & 20 & 1 & 95\% \\
\midrule
\textbf{Total}            & \textbf{55} & \textbf{37} & \textbf{62} & \textbf{154} & \textbf{12} & \textbf{92\%} \\
\bottomrule
\end{tabular}
\end{table}

\subsubsection{Control Granularity Sensitivity}
\label{subsec:granularity}

Consolidation of this kind invites an obvious question. A catalog of twelve controls has to justify twelve, and a reader is entitled to ask whether a different abstraction level would change the conclusions. The engine answers it by re-clustering the same 154 obligations at a coarser and a finer level and recomputing the structural properties at each. Table~\ref{tab:granularity} reports the three levels. Loosening the intent test so that any two obligations sharing a governance object merge, regardless of the required action, yields eight controls, one per governance domain. Tightening it so that obligations merge only within a single framework yields thirty-six.

Finer granularity fails decisively, and it fails on the property the crosswalk exists to produce. Tri-framework convergence falls from 100\% to zero at the per-framework level, because every control then draws from exactly one source. The catalog becomes three parallel control libraries carrying the overhead of a unified vocabulary without the benefit, which is the condition the crosswalk exists to remove. Consolidation drops to 76.6\% at the same time.

Coarser granularity gives a less comfortable result, and the honest report is that no structural metric distinguishes it from the published level. Tier-activation signatures, artifact bindings, and tri-framework convergence are identical at eight controls and at twelve. Consolidation rises to 94.8\%, which follows definitionally from a smaller control count and not from better harmonization. Results reported in this paper would be unchanged had the catalog stopped at the domain level.

What separates the two levels is not visible to the current metric set. Every domain holding more than one control holds controls that run on different lifecycle phases. Risk assessment spans design, monitoring, and change, while risk treatment spans design, deployment, and change. Oversight design spans design and build, while override and intervention spans design, deployment, and monitoring. Data quality and bias detection diverge, as do accuracy validation and robustness testing. A domain-level merge erases four such distinctions, each separating activities with different implementation owners and different points of execution. Twelve controls preserve them.

The conclusion is bounded accordingly. Granularity between the domain level and the published level sits inside a band where the reported structural results are invariant, and the choice of twelve rests on lifecycle and action separability and not on any metric advantage. Below that band the catalog would conflate distinct required actions. Above it, harmonization degrades measurably. A weighted metric sensitive to lifecycle placement would discriminate inside the band, and building one requires the stakeholder-validated weights that Section~\ref{subsec:coverage} places outside current scope.

\begin{table}[!t]
\centering
\caption{Control granularity sensitivity. The same 154 source obligations re-clustered at three abstraction levels. Tri-framework convergence is the share of controls drawing on all three sources.}
\label{tab:granularity}
\footnotesize
\renewcommand{\arraystretch}{1.15}
\begin{tabular}{@{} L{2.4cm} c c c c @{}}
\toprule
\textbf{Clustering level} & \textbf{Controls} & \textbf{Consolid.} & \textbf{Tri-fw} & \textbf{Tier sigs.} \\
\midrule
L1 coarse, one per domain & 8 & 94.8\% & 100\% & 4 \\
L2 published, object $\times$ action & 12 & 92.2\% & 100\% & 4 \\
L3 fine, one per framework & 36 & 76.6\% & 0\% & 4 \\
\bottomrule
\end{tabular}
\end{table}

\subsubsection{Cross-Tier Dependencies}
 
The engine identifies five cross-tier evidence chains in three-tier deployments, summarized below. Each chain represents a sequence of controls and artifacts that spans organizational boundaries and cannot be represented under siloed compliance.

\begin{enumerate}[label=\textbf{Chain~\arabic*.},leftmargin=3.5em,itemsep=1pt]
\item \textbf{Incident response escalation} (T2$\rightarrow$T3$\rightarrow$T3$\rightarrow$T1/T2): Edge anomaly trigger (UC-12) $\rightarrow$ cloud risk update (UC-01, UC-02) $\rightarrow$ controlled model change with documentation update (UC-07) $\rightarrow$ redeployment validation (UC-09). This is the longest chain at four UC steps.
\item \textbf{Model update governance} (T3$\rightarrow$T3$\rightarrow$T2$\rightarrow$T1): Cloud model retraining and supplier verification (UC-09, UC-11) $\rightarrow$ OTA distribution with documentation (UC-07) $\rightarrow$ edge deployment validation (UC-10) $\rightarrow$ vehicle accuracy verification (UC-09).
\item \textbf{Data quality pipeline} (T2$\rightarrow$T3$\rightarrow$T3$\rightarrow$T2): Edge data collection and quality checks (UC-03) $\rightarrow$ cloud bias analysis (UC-04) $\rightarrow$ cloud risk assessment (UC-01) $\rightarrow$ edge remediation and robustness re-testing (UC-10).
\item \textbf{Oversight coordination} (T1$\rightarrow$T2$\rightarrow$T3): Vehicle oversight design (UC-05) $\rightarrow$ edge override protocol alignment (UC-06) $\rightarrow$ cloud documentation and risk treatment update (UC-07, UC-02).
\item \textbf{Supplier propagation} (T3$\rightarrow$T3$\rightarrow$T2$\rightarrow$T1): Cloud supplier assessment (UC-11) $\rightarrow$ documentation update (UC-07) $\rightarrow$ edge deployment validation (UC-09) $\rightarrow$ vehicle security testing (UC-10).
\end{enumerate}

Rural, with a single tier, has zero cross-tier chains. This is not a deficiency. It is the framework reporting, correctly, that a single-tier deployment has no cross-organizational evidence coordination to govern.
 
These chains are the structures that siloed compliance cannot represent. A team operating under ISO/IEC~42001 alone would track the incident through its management-system audit trail. A team preparing EU AI Act evidence would document the same incident through Annex~IV technical documentation. A team following NIST AI RMF would log it under MANAGE~4 incident response. Three parallel records of the same event, with no guarantee of consistency and no shared trace identifiers. The cross-tier dependency computation makes this coordination requirement visible and auditable.

\subsection{Comparison with Related Tool-Backed Approaches}\label{subsec:tool_comparison}
 
Three recent tool-backed approaches provide relevant comparison points. MARISMA~\cite{marisma2025} validates a cybersecurity risk management framework through the eMARISMA cloud platform across multiple sectors and countries, establishing the gold standard for tool-backed validation at CSI. MARISMA addresses a single standards family (ISO/IEC~27000 and 31000) in a single risk domain, while UGAF-ITS harmonizes three heterogeneous instruments across eight governance domains in a specific application domain. The validation approaches share a common structure: software tool, case study instantiation, quantitative metrics, and open reporting of scope boundaries.
 
PASTA~\cite{pasta2026} uses large language models to evaluate AI systems against five policies simultaneously, achieving Spearman $\rho \geq 0.626$ correlation with expert judgments. PASTA treats regulations individually without harmonizing them into unified controls, and it targets runtime compliance assessment and not design-time governance architecture. Sovrano~et~al.~\cite{sovrano2025} validate AI-assisted EU AI Act technical documentation through expert comparison, achieving statistically significant agreement. Their contribution targets a single framework with a different automation strategy (document generation and not control consolidation). Neither PASTA nor Sovrano addresses distributed system architectures or tier-aware evidence allocation. Tri-framework scope, distributed ITS application, and tier-aware governance allocation distinguish UGAF-ITS from the tool-backed approaches surveyed here.

\subsection{Threats to Validity}\label{subsec:threats}
 
Five validity threats bound the conclusions.
 
\textbf{Construct validity.} Coverage (M1) counts mapped obligations, not mapping quality. A covered obligation may vary in legal strictness across frameworks; the crosswalk tie-break rules (Section~\ref{sec:crosswalk}) mitigate but do not eliminate this ambiguity. Instantiation depth uses a weighted composite that reflects the authors' judgment about factor importance; alternative weights would shift the scores without changing the ordinal ranking across scenarios.
 
\textbf{Internal validity.} The siloed baseline is enumerated document by document in \ref{app:register} and computed by the engine, which removes the earlier reliance on an unexplained document count. Enumeration is not measurement. Each entry encodes a judgment about what a siloed programme would produce for that clause, and a practitioner running a real ISO, EU, and NIST programme in parallel might produce a different register. Alternative registers would change M2 and M3 magnitudes, so the register is released as source instead of as a reported total.
 
\textbf{External validity.} The four scenarios reflect ITS archetypes. Highway corridors with minimal edge logic, rural deployments with connectivity constraints, and multi-modal transit systems may shift which tiers dominate governance. The engine and scenarios are open-source; extending to additional archetypes requires only a YAML descriptor.
 
\textbf{Crosswalk subjectivity.} The coding protocol was applied by the research team without independent inter-rater reliability measurement. The codebook, tie-break rules, and decision log are publicly available in the repository, so an independent team can reproduce the process. That said, reproducibility is not the same as agreement. We plan a formal inter-rater study as the immediate next step: two independent governance practitioners will code a stratified sample of 30 obligations (10 per source framework, spread across governance domains) using the published codebook. We will measure Cohen's $\kappa$ on domain assignment, obligation type classification, and tier applicability. The target is $\kappa \geq 0.70$ (substantial agreement). If the codebook produces unstable codings across independent raters, the crosswalk needs revision regardless of how well the engine runs.

\textbf{Temporal validity.} The crosswalk reflects the three frameworks as published in early 2026. All three are moving targets. The EU AI Act's implementing acts are still being developed. Harmonised standards including prEN~18286 are in public enquiry. NIST continues to publish companion resources (the Generative AI Profile, NIST AI 600-1). The governance engine's knowledge base supports incremental obligation insertion, so adding or modifying individual obligations and their trace links does not require reconstructing the entire catalog. What we have not measured is the effort a crosswalk maintenance cycle actually takes. That estimate requires operational experience with at least one standards revision.

\section{Discussion}
\label{sec:discussion}

The multi-scenario evaluation indicates that governance harmonization for distributed ITS is structurally feasible at the level the design permits, namely 91.7\% average framework coverage across three-tier deployments, a 45.9\% reduction in documentation volume, and 80\% of artifacts reused across all three frameworks simultaneously. These numbers matter because ITS operators face near-term multi-framework pressure (including the EU AI Act high-risk deadline in August~2026) while continuing to meet safety and cybersecurity expectations already embedded in transportation engineering practice. The results hold across architecturally distinct deployment scenarios and contract predictably when tiers are removed. Reduction stays between 25\% and 50\% across every archetype, computed against an enumerated siloed register, not an assumed document count.

The question is what these results mean in practice and where the boundaries lie.

\subsection{Comparison with Alternative Approaches}
\label{subsec:comparison}

ITS programs typically handle multi-framework pressure in one of three ways: siloed compliance (separate control libraries and audit preparation per framework), framework-first mapping (one framework as anchor with others mapped into it), or control library consolidation (unified catalog without tier allocation). Table~\ref{tab:approach_comparison} compares these strategies against UGAF-ITS across six criteria. Siloed approaches preserve framework fidelity but duplicate evidence and force post-incident reconstruction of parallel trails~\cite{Batool2025AIGovernanceSLR}. Framework-first mapping creates uneven treatment for requirements that do not align with the anchor~\cite{hernandez2025open}. Consolidation stops at control naming and leaves the key ITS question unanswered: which tier owns enforcement~\cite{Avianti2025GRCBibliometric}.

UGAF-ITS adds what these strategies lack. Twelve unified controls are attached to vehicle, edge/RSU, and cloud tiers based on enforceability and observability, and the evidence backbone enables reuse without losing the bidirectional traceability that framework-specific audits require. Section~\ref{subsec:tool_comparison} further positions UGAF-ITS against recent tool-backed approaches (MARISMA, PASTA, Sovrano~et~al.), where the distinguishing contribution is the combination of tri-framework scope, distributed ITS application, and tier-aware governance allocation.

\begin{table}[!t]
\caption{Governance approaches for multi-framework ITS compliance.}
\label{tab:approach_comparison}
\centering
\footnotesize
\setlength{\tabcolsep}{3pt}
\renewcommand{\arraystretch}{1.12}
\begin{tabularx}{\columnwidth}{@{} l X X X X @{}}
\toprule
\textbf{Criterion} & \textbf{Siloed} & \textbf{Fw-first} & \textbf{Control lib.} & \textbf{UGAF-ITS} \\
\midrule
Fw coverage          & High & Part.--High & High    & High \\
Control consolidation & None & Partial    & Yes     & Yes \\
Tier allocation       & None & None       & None    & \textbf{Yes} \\
Evidence reuse        & None & Partial    & Partial & \textbf{Yes} \\
ITS specificity       & No   & No         & No      & \textbf{Yes} \\
Traceability          & Per-fw & Partial  & Partial & \textbf{Bidirect.} \\
\bottomrule
\end{tabularx}
\end{table}

\subsection{What UGAF-ITS Enables}
\label{subsec:discussion_enables}

Beyond structural differentiation from alternative strategies, UGAF-ITS changes what governance produces and how teams interact with it.

\textbf{From compliance burden to governance asset.} Under traditional approaches, multi-framework compliance remains overhead: teams generate documentation for auditors, then set it aside until the next assessment cycle. UGAF-ITS inverts that relationship. The unified control vocabulary becomes an operational tool that teams use to scope work, track evidence completeness, and communicate across organizational boundaries. One vocabulary. Three frameworks. Twelve controls allocated across vehicle, edge, and cloud tiers, while traceability to source identifiers remains intact.

The evidence backbone turns artifacts into living objects and not a compliance archive. Version-controlled artifacts capture operational decisions, incident response actions, and change rationale in formats that serve engineering workflows and audit questioning. The cross-tier incident chain described in Section~\ref{subsec:cross_tier} illustrates the mechanism: evidence produced during incident response (decision logs, configuration baselines, mitigation records, closure checks) satisfies UC-12 while also supporting UC-07 and UC-01/UC-02. Operational execution and compliance evidence become the same activity, reducing duplication and improving post-incident accountability.

\textbf{The UGAF-ITS principle.} Governance follows architecture. Controls attach to tiers, not to frameworks. Evidence is produced where behavior executes, not where compliance is reported.

\subsection{Integration with Safety Assurance and V2X Practice}
\label{subsec:discussion_interop}

UGAF-ITS integrates with established transportation assurance practice. The integration points are concrete and map to artifacts that safety and V2X teams already recognize.

\textbf{Functional safety (ISO~26262).} UC-01 extends hazard analysis with AI-specific failure modes, such as data drift, distribution shift, and adversarial perturbation. UC-09 complements software-in-the-loop and hardware-in-the-loop verification with AI test coverage expectations tied to the operational design domain. UC-07 supplies AI-specific evidence packages that fit into safety case argumentation and change governance without introducing a parallel documentation system.

\textbf{Safety of the intended functionality (SOTIF) (ISO~21448).} UC-03, UC-04, and UC-10 operationalize concerns that Safety of the Intended Functionality addresses: performance limitations, foreseeable misuse, and safety-relevant behavior under uncertainty. Data-quality reports, bias and performance reports, and robustness evidence populate SOTIF triggering condition analysis and validation strategy records, while preserving trace links to governance obligations.

\textbf{V2X practice.} V2X message integrity depends on trust in the systems that generate and transform intersection state at the roadside tier. UGAF-ITS does not prescribe message formats, but it makes message-generating AI auditable. UC-08 supports confidence signaling and operator-facing interpretability appropriate to the deployment context, while UC-10 requires testing against spoofing, injection, and replay conditions relevant to V2X threat models. The evidence backbone supplies the documentation trail that public key infrastructure enrollment, incident investigations, and controlled rollout workflows increasingly require.

\subsection{Scope and Limitations}
\label{subsec:discussion_limits}

Three boundaries constrain the current contribution beyond the validity threats addressed in Section~\ref{subsec:threats}.

\textbf{Deployment coverage.} The four scenarios span the urban--rural and three-tier--single-tier spectrum, but they do not cover all ITS archetypes. Multi-modal transit hubs, highway corridors with minimal edge logic, and deployments under connectivity constraints may shift which tiers dominate governance. The open-source engine and YAML-based scenario descriptors are designed to support such extensions with minimal effort.

\textbf{Evidence exchange assumptions.} Multi-vendor ITS delivery depends on evidence exchange across contractual boundaries. UGAF-ITS reduces ambiguity by allocating controls to tiers, but the approach still assumes that vendors can provide minimally sufficient evidence objects (hashes, signed summaries, configuration baselines) without disclosing protected internals.

\textbf{Framework evolution.} ISO standards evolve, the EU AI Act will be complemented by guidance and harmonized standards, and national policies continue to emerge. Two recent developments are particularly relevant: prEN~18286~\cite{pren18286}, the first harmonised standard for AI under the EU AI Act, entered public enquiry in October~2025 and targets Article~17 quality management requirements; and ISO/IEC~42005:2025~\cite{iso42005} introduces AI impact assessment guidance that complements ISO/IEC~42001's management-system structure.

A preliminary analysis suggests prEN~18286 would not require new unified controls. Its quality management system requirements map to UC-07 (System Documentation), which already uses EU AI Act Annex~IV as the minimum content template. Its risk management provisions align with UC-01 and UC-02, where the strictest lifecycle-continuous requirements from Article~9 already set the baseline. Its monitoring and corrective action requirements align with UC-12 (Incident Response). The integration path is therefore incremental: add prEN~18286 clause references as additional trace links within existing UCs, and where the standard introduces sub-requirements more specific than what is currently encoded, add those as framework-specific extensions. No new UCs. No crosswalk reconstruction. The engine's knowledge base supports this kind of incremental insertion. What remains untested is the actual effort required to validate semantic alignment for a standard that is still in public enquiry.

UGAF-ITS preserves traceability to source identifiers, but long-term stability depends on maintaining crosswalk updates and validating that unified controls remain semantically aligned as obligations shift and new standards enter the ecosystem.

\subsection{Implications for Practice}
\label{subsec:implications_practice}

Three practical implications follow for ITS teams that must converge on multi-framework governance.

\textbf{Begin with tier mapping, then map obligations.} Mapping system architecture to vehicle, edge/RSU, and cloud tiers clarifies where controls can be enforced and which party can produce defensible evidence. Unified controls become easier to assign when tier boundaries anchor accountability.

\textbf{Treat the Technical File as infrastructure.} Technical File operations (version control, trace link management, and change notification) determine whether artifact reuse is achievable. Documentation reduction depends on shared infrastructure more than on clause-by-clause writing.

\textbf{Design incident response as evidence production.} Incident workflows should capture the artifacts that UC-12, UC-07, and UC-01/UC-02 require, so post-incident governance remains navigable across tiers without reconstruction.

A phased adoption fits most programs approaching the EU AI Act timeline: map the deployed architecture to tiers, assess current artifacts against the unified control catalog, establish Technical File infrastructure, then prioritize gap closure for high-risk AI components and cross-tier incident pathways. Table~\ref{tab:ugaf_its_roadmap} presents a phased adoption roadmap.

\begin{table*}[!t]
\centering
\caption{Phased UGAF-ITS adoption roadmap for ITS programs approaching the EU AI Act August~2026 deadline.}
\label{tab:ugaf_its_roadmap}
\footnotesize
\renewcommand{\arraystretch}{1.15}
\setlength{\tabcolsep}{5pt}
\begin{tabular}{@{} L{3.0cm} L{7.5cm} L{2.0cm} L{3.8cm} @{}}
\toprule
\textbf{Phase} & \textbf{Activity} & \textbf{Timeline} & \textbf{Key Output} \\
\midrule
1. Architecture mapping &
Map deployed AI components to vehicle, edge, and cloud tiers; identify ownership boundaries &
Weeks 1--2 &
Tier map and stakeholder RACI \\[3pt]

2. Gap assessment &
Assess current artifacts against the 12 unified controls (Table~\ref{tab:unified_controls}) &
Weeks 3--4 &
Gap register per tier \\[3pt]

3. Evidence infrastructure &
Establish the Technical File as shared infrastructure with version control and trace links &
Weeks 5--8 &
Technical File skeleton and trace map \\[3pt]

4. Priority closure &
Close gaps for high-risk AI components and cross-tier incident pathways &
Weeks 9--16 &
Populated evidence backbone \\[3pt]

5. Audit readiness &
Conduct an internal walkthrough simulating framework-specific audit questions &
Week 17+ &
Validated audit package \\
\bottomrule
\end{tabular}
\end{table*}

\subsection{Societal Implications}
\label{subsec:societal}

Governance fragmentation carries direct societal consequences. When incident accountability is unclear across tiers, affected parties face opaque redress pathways. UGAF-ITS addresses this by making accountability architecturally explicit: each unified control attaches to the tier where behavior executes, so the responsible party is identifiable by design and not determined through post-incident negotiation.

\subsection{Implications for Research}
\label{subsec:implications_research}

The released artifacts are designed for reuse, not for inspection alone. Adding a fourth instrument requires extracting its obligations to atoms, running them through the three alignment tests against the existing catalog, and attaching the results as trace links or framework-specific extensions. No unified control need be reconstructed, because the catalog is keyed to governance objects and required actions, not to sources. Retargeting the framework to another distributed domain requires replacing the tier definitions and the artifact set while the crosswalk, the decision procedure, and the siloed register construction transfer unchanged. Both operations are exercised by the scenario descriptors and the register already in the repository.

UGAF-ITS opens several research directions beyond scenario-based validation. Governance automation remains a natural next step because structured crosswalks and artifact schemas support continuous compliance checks, auto-population of evidence templates, and drift detection against validated configurations. Multi-jurisdictional extension will test scalability as ITS deployments increasingly face additional national frameworks with divergent requirements. Governance usability also deserves direct study because traceability quality depends on how practitioners navigate controls, maintain links, and respond under time pressure.

Empirical validation requires methods that quantify mapping stability and operational maintainability. Inter-rater agreement studies using stratified samples of obligations can measure crosswalk consistency, while sensitivity analysis can bound consolidation metrics under alternative baselines. Shadow audits on mature open-source ITS components and pipelines can test whether tier-aware evidence exposes governance gaps that siloed approaches miss. Replication packages that include the crosswalk, artifact templates, and the open-source governance engine strengthen audit-defensibility claims and support independent reuse.

\section{Conclusion}\label{sec:conclusion}
 
Governance follows architecture. When systems distribute, governance must distribute with them. UGAF-ITS makes that distribution auditable.
 
This paper introduced a standards harmonization framework that consolidates 154 obligations from ISO/IEC~42001, the EU AI Act, and the NIST AI RMF into 12 unified controls. Those controls are allocated to vehicle, edge, and cloud tiers based on where behavior executes and evidence is observable. Twenty reusable evidence artifacts bind them into a single audit package across all three frameworks. The crosswalk methodology is reproducible. The tier model is architecturally grounded. The evidence backbone reduces documentation volume without weakening traceability.
 
An open-source governance engine computed these properties across four ITS deployment scenarios that vary from a 12-component, three-tier, multi-stakeholder urban intersection to a 2-component, single-tier rural installation. Three-tier deployments achieve 91.7\% average scoped framework coverage, 45.9\% evidence reduction, complete bidirectional traceability, and 80\% of artifacts serving all three frameworks simultaneously. Partial deployments contract predictably. Consolidation holds between 25\% and 50\% across all four archetypes without rising with tier count. The consolidation analysis confirms that every governance domain draws from all three source frameworks, granularity sensitivity shows the reported properties to be invariant across an eight-to-twelve control band, and the cross-tier dependency computation surfaces the evidence coordination chains that siloed compliance approaches cannot represent.
 
Three limitations bound the current contribution. The crosswalk coding protocol was applied by the research team without independent inter-rater reliability measurement, the deployment scenarios are synthetic architectural descriptors and not field evaluations with operational ITS programs, and the siloed baseline register encodes a judgment about siloed practice that no operational programme has yet confirmed. Future work will address all three: inter-rater agreement studies using stratified samples of obligations, shadow audits on operational ITS pipelines to test whether tier-aware evidence allocation exposes governance gaps that siloed approaches miss. Multi-jurisdictional extension will test scalability as deployments face additional national frameworks beyond the three harmonized here.
 
The framework is designed to generalize beyond ITS. Any distributed AI system with clear tier boundaries and multi-party ownership faces the same coordination failure modes. The crosswalk methodology, tier-aware allocation, and evidence backbone are designed to transfer; only tier definitions and domain-specific interpretations require adaptation. Validation in non-ITS domains, such as smart grids, industrial IoT, and connected medical devices, remains future work.

\section*{Data Availability}\label{sec:data_availability}

The UGAF-ITS Governance Engine source code, deployment scenario descriptors, knowledge base encoding all 154 source obligations and 12 unified controls, the siloed baseline register of \ref{app:register}, and the complete results reported in this paper are available at \url{https://doi.org/10.5281/zenodo.19339071} under the MIT license. That identifier is a concept DOI and resolves to the most recent release; the results reported here were produced with version~1.2.0. Scenario results and the HTML dashboard reproduce with \texttt{python run\_validation.py}. The granularity sensitivity analysis of Section~\ref{subsec:granularity} and the tier-scoped baseline derivation reproduce with \texttt{python run\_granularity.py}. The result figures reproduce with \texttt{python make\_figures.py}.

\section*{Acknowledgments}
 
The authors acknowledge the support of the Higher Colleges of Technology, United Arab Emirates.

%======================================================================

\appendix

\section{Siloed Baseline Register}
\label{app:register}

Table~\ref{tab:register} enumerates the documentation set a siloed compliance programme produces when each framework is satisfied independently. Each entry records the producing framework, the clause or article that requires the document, the tiers at which it is generated, and the unified controls it would evidence. Entries marked as regulatory workflow evidence no unified control, because the activity sits outside the technical control perimeter, and a siloed programme produces them regardless.

The governance engine computes every reported reduction figure from this register. A document counts for a deployment when at least one producing tier is active and, where the document evidences unified controls, at least one of those controls is activated. The same rule governs the unified artifact set, so neither side of the comparison can be scoped more generously than the other.

\begin{table*}[!t]
\centering
\caption{Siloed baseline register. Thirty-seven documents across the three frameworks, with source clause, producing tiers, and evidenced unified controls.}
\label{tab:register}
\footnotesize
\renewcommand{\arraystretch}{1.1}
\begin{tabular}{@{} l L{8.3cm} L{2.9cm} l L{2.6cm} @{}}
\toprule
\textbf{ID} & \textbf{Document} & \textbf{Source} & \textbf{Tiers} & \textbf{Evidences} \\
\midrule
\multicolumn{5}{@{}l}{\textit{ISO/IEC 42001 (15 documents)}} \\
ISO-D01 & AI policy & Cl.\,5.2 & T3 & UC-01 \\
ISO-D02 & AIMS scope statement & Cl.\,4.3 & T3 & UC-01 \\
ISO-D03 & Roles and responsibilities record & Cl.\,5.3 & T3 & UC-01 \\
ISO-D04 & AI risk assessment results & Cl.\,6.1, 8.2 & T1--T3 & UC-01 \\
ISO-D05 & Risk treatment plan and Statement of Applicability & Cl.\,6.1, 8.3 & T3 & UC-02 \\
ISO-D06 & AI objectives and planning record & Cl.\,6.2 & T3 & UC-02 \\
ISO-D07 & Competence records & Cl.\,7.2 & T3 & UC-05 \\
ISO-D08 & Communication records & Cl.\,7.4 & T3 & UC-08 \\
ISO-D09 & Operational planning and control records & Cl.\,8.1 & T2, T3 & UC-02 \\
ISO-D10 & AI system impact assessment & A.5 & T3 & UC-01 \\
ISO-D11 & Data management records & A.7 & T2, T3 & UC-03, UC-04 \\
ISO-D12 & AI system information for users & A.13, A.14 & T1, T2 & UC-08 \\
ISO-D13 & Human oversight records & A.15 & T1, T2 & UC-05, UC-06 \\
ISO-D14 & Internal audit programme and results & Cl.\,9.2 & T3 & Regulatory workflow \\
ISO-D15 & Management review records & Cl.\,9.3 & T3 & Regulatory workflow \\
\midrule
\multicolumn{5}{@{}l}{\textit{EU AI Act (12 documents)}} \\
EU-D01 & Annex~IV general system description & Art.\,11, Annex~IV(1) & T3 & UC-07 \\
EU-D02 & Annex~IV design specification and architecture & Annex~IV(2) & T3 & UC-07 \\
EU-D03 & Training and validation data description & Art.\,10, Annex~IV(2d) & T2, T3 & UC-03, UC-04 \\
EU-D04 & Validation and testing procedures and results & Art.\,15, Annex~IV(2g) & T1--T3 & UC-09 \\
EU-D05 & Risk management system documentation & Art.\,9 & T3 & UC-01, UC-02 \\
EU-D06 & Instructions for use & Art.\,13 & T1 & UC-08 \\
EU-D07 & Human oversight measures description & Art.\,14 & T1, T2 & UC-05, UC-06 \\
EU-D08 & Automatically generated logs & Art.\,12 & T1, T2 & UC-12 \\
EU-D09 & Quality management system documentation & Art.\,17 & T3 & UC-07 \\
EU-D10 & Cybersecurity and robustness evidence & Art.\,15(4), 15(5) & T1--T3 & UC-10 \\
EU-D11 & Post-market monitoring plan and reports & Art.\,72 & T2, T3 & UC-12 \\
EU-D12 & Declaration of conformity and registration record & Art.\,47, 49 & T3 & Regulatory workflow \\
\midrule
\multicolumn{5}{@{}l}{\textit{NIST AI RMF (10 documents)}} \\
NIST-D01 & AI RMF profile and context mapping record & MAP 1, MAP 5 & T3 & UC-01 \\
NIST-D02 & Risk measurement plan & MEASURE 1.1 & T3 & UC-02 \\
NIST-D03 & TEVV results record & MEASURE 2.5 & T1--T3 & UC-09 \\
NIST-D04 & Bias and fairness assessment & MEASURE 2.11 & T2, T3 & UC-04 \\
NIST-D05 & Transparency and explainability record & MEASURE 2.9 & T1, T2 & UC-08 \\
NIST-D06 & Monitoring and alerting plan & MANAGE 4.1 & T2, T3 & UC-12 \\
NIST-D07 & Incident response record & MANAGE 4.1, 4.2 & T1--T3 & UC-12 \\
NIST-D08 & Third-party component register & GOVERN 6.1 & T3 & UC-11 \\
NIST-D09 & Governance roles and accountability record & GOVERN 2.1 & T3 & Regulatory workflow \\
NIST-D10 & Change management and decommissioning record & MANAGE 4.3 & T3 & UC-07 \\
\bottomrule
\end{tabular}
\end{table*}

\bibliographystyle{elsarticle-num}
\bibliography{references}
%======================================================================

\end{document}